\documentclass[sigconf]{acmart}
\renewcommand\footnotetextcopyrightpermission[1]{}

\usepackage{kotex}
\usepackage{listings}
\usepackage{graphicx}
\usepackage{subfigure}
\usepackage[ruled,linesnumbered,vlined]{algorithm2e}
\usepackage[normalem]{ulem}
\usepackage{multirow}
\usepackage{xspace}
\usepackage{enumitem}
\usepackage{xr}

\newcommand{\de}[2]{\{#1\} $\rightarrow$ \{#2\}}
\newcommand{\node}[1]{$\langle$#1$\rangle$}
\newcommand{\wc}{\textsf{wc}\xspace}
\newcommand{\tcas}{\textsf{tcas}\xspace}
\newcommand{\sch}{\textsf{schedule}\xspace}
\newcommand{\scht}{\textsf{schedule2}\xspace}
\newcommand{\tot}{\textsf{totinfo}\xspace}
\newcommand{\prt}{\textsf{printtokens}\xspace}
\newcommand{\prtt}{\textsf{printtokens2}\xspace}
\newcommand{\rep}{\textsf{replace}\xspace}
\makeatletter
\newcommand\footnoteref[1]{\protected@xdef\@thefnmark{\ref{#1}}\@footnotemark}
\makeatother

\begin{document}
  \title{Causal Program Dependence Analysis}

  \author{Seongmin Lee}
  \affiliation{
    \institution{KAIST}
    \streetaddress{291 Daehak Ro, Yuseong Gu}
    \city{Daejeon}
    \country{Republic of Korea}
    \postcode{34141}
  }
  \email{bohrok@kaist.ac.kr}

  \author{Dave Binkley}
  \affiliation{
    \institution{Loyola University Baltimore}
    \streetaddress{4501 North Charles Street}
    \city{Baltimore}
    \country{USA}
    \postcode{MD 21210}
  }
  \email{binkley@cs.loyola.edu}

  \author{Robert Feldt}
  \affiliation{
    \institution{Chalmers University of Technology}
    \streetaddress{Chalmersplatsen 4}
    \city{Göteborg}
    \country{Sweden}
    \postcode{412 96}
  }
  \email{robert.feldt@chalmers.se}

  \author{Nicolas Gold}
  \affiliation{
    \institution{University College London}
    \streetaddress{Gower St}
    \city{London}
    \country{UK}
    \postcode{WC1E 6BT}
  }
  \email{n.gold@ucl.ac.uk}

  \author{Shin Yoo}
  \affiliation{
    \institution{KAIST}
    \streetaddress{291 Daehak Ro, Yuseong Gu}
    \city{Daejeon}
    \country{Republic of Korea}
    \postcode{34141}
  }
  \email{shin.yoo@kaist.ac.kr}

\begin{abstract}

We introduce Causal Program Dependence Analysis (CPDA), a dynamic 
dependence analysis that applies causal inference to model the strength of 
program dependence relations in a continuous space.
CPDA observes 
the association between program elements by constructing and executing
modified versions of a program.
One advantage of CPDA is that this construction requires only light-weight
parsing rather than sophisticated static analysis. 
The result is a collection of observations based on how often a change in the value
produced by a mutated program element affects the behavior of other elements. 
From this set of observations, CPDA discovers a causal structure capturing the
causal (i.e., dependence) relation between program elements. 
Qualitative evaluation 
finds that CPDA concisely expresses key dependence relationships between program elements. 
As an example application, we apply CPDA to the problem of fault localization.
Using minimal test suites, our approach can rank twice as many faults  compared to SBFL.

\end{abstract}

\maketitle
\pagestyle{plain}

\section{Introduction}

Program dependence analysis is a fundamental task in software engineering. 
In addition to being the foundation of program comprehension~\cite{
Zhifeng-Yu:2001aa}, program dependence analysis underlies various software 
engineering tasks, including software testing~\cite{Binkley:1997aa}, 
debugging~\cite{Jiang:2017aa,Lee:2020aa}, refactoring~\cite{Ettinger:2004aa}, 
maintenance~\cite{Gallagher:1991aa}, and security~\cite{Karim:2018aa}. It 
often supports these tasks by reducing the number of program elements that must
be considered.

Each component of a program depends, to some degree, on the other components of
the program.
Some of these dependence relations are strong, i.e. a change in one component almost surely changes the value of another, while others are weaker.
Traditional static dependence analysis attempts to safely approximate \emph{all} possible 
dependence relations. This can become quite involved, and will return many relations.
A good example would be pointer analysis, which is not only computationally expensive but also prone to produce a large number of dependence relations including many false positives.
Furthermore, static analysis typically does not attempt to estimate the strength of
a dependence.  Thus its output is a large number of equally-weighted dependencies, which tends to degrade the ability of a developer to work with the code based on the results of the analysis.
A few dynamic dependence analysis approaches have modeled dependence
strengths~\cite{Baah:2010aa,Yu:2017ab}, but they are typically built
on top of (expensive) static dependence analysis (e.g., the construction of a
Program Dependence Graph), and remain vulnerable to confounding bias due to 
their formulation based on conditional probability, which captures mere association, 
rather than causation~\cite{Baah:2010aa,Yu:2017ab,Lee:2019aa}.

This paper proposes \emph{Causal Program Dependence Analysis (CPDA)}, a 
framework that analyzes the degree (or strength) of the dependence between 
program elements. Given a set of executions, CPDA makes use of associations 
between observed behavior of program elements to discover a \emph{causal structure}.
By applying techniques for causal inference~\cite{Pearl:2009aa,Pearl:2009causalOverview},
over this causal structure, 
CPDA then produces two measures of program dependence:
\emph{causal dependence} and \emph{direct dependence}.
Causal dependence aggregates the effect that a change at a program element
has on the behavior of other program elements. 
A direct dependence is the effect of one program element on another, excluding
any effects that pass through other elements. 

One advantage of CPDA is that it does not require extensive ``heavy 
weight'' static analysis. Instead, only light-weight parsing of the source 
code, or light-weight instrumentation of a binary, is required. 
Essential points of interest in the code are annotated so we can 
introduce simple mutations to the state, and observe their effects
using a set of test inputs. 
An additional benefit is thus that our approach can be applied to 
heterogeneous 
systems built from multiple languages, as long as a parser
(or method for instrumenting binary representations) is available.

To evaluate CPDA, we first build a \emph{Causal Program Dependence Model
(CPDM)}, and then introduce and study its use for \emph{Causal Dependence based Fault
Localization (CDFL)}. 
The CPDM is a graph in which we annotate the direct dependence on the causal 
structure to represent a program's dependences. We investigate how 
well the (strength of the) direct dependence captures the program's semantics. 
We also consider how characteristics of the test suite impacts the resulting CPDM.

To further study the utility of CPDA, we propose \emph{Causal Dependence based
Fault Localization (CDFL)}, which calculates the suspiciousness of program 
elements in light of failing test cases, based on CPDA.
We evaluate CDFL using the Siemens suite, a well-known benchmark for 
testing and debugging. 
Compared to Spectrum based Fault Localization (SBFL)~\cite{Naish:2011fk,Xie:2013uq},
our empirical results CDFL can rank 166\% more actually faulty program 
elements at the top.

\begin{figure*}[ht]
  \includegraphics[width=\textwidth,trim={87 448 107 361},clip]{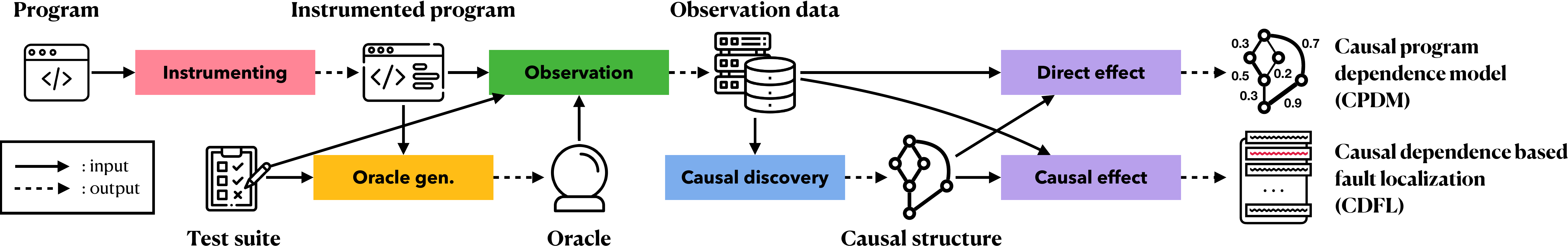}
  \Description{Framework of Causal Program Dependence Analysis}
  \caption{Framework of CPDA}
  \label{fig:framework}
\end{figure*}

The main contributions of this paper include the following:

\begin{itemize}
\item We propose Causal Program Dependence Analysis. This dependence analysis
framework, based on causal inference, supports the quantification of 
dependence strength among program elements.

\item We present an illustrative example 
where quantifiable program dependence enables us to visualize program semantics 
and the effect of different inputs on them.

\item To illustrate the utility of CPDA, we introduce Causal Dependence based 
Fault Localization (CDFL). Our empirical evaluation of CDFL shows that it 
outperforms existing fault localization techniques, especially in challenging
cases that other techniques find difficult.
\end{itemize}

The rest of this paper is organized as follows. We first describe CPDA, 
starting with an introduction of causal inference in Section~\ref{sec:cpda}. 
Section~\ref{sec:experimental_setup} describes the design of our empirical 
evaluation, the results of which are presented in Section~\ref{sec:results}.
Section~\ref{sec:discussion} discuss the remaining challenges in CPDA and 
presents future work. Threats to validity are discussed in 
Section~\ref{sec:threats}. Finally, we review related work in 
Section~\ref{sec:relatedwork}, and conclude in Section~\ref{sec:conclusion}.

\section{Causal Program Dependence Analysis (CPDA)}
\label{sec:cpda}

This section first introduces causal inference in general, and 
subsequently presents a detailed explanation of our framework, Causal Program 
Dependence Analysis (CPDA).

\subsection{Causal Inference}

Causal inference is a fundamental, mathematical theory for analyzing which factors cause one or more effects~\cite{Pearl:2009aa,Pearl:2009causalOverview}. 
While statistical methods focus on identifying and modeling associations, causal analysis adds ways of studying which factors actually precede and leads to (causes) changes in others, and how large their effects are.
Such an analysis can provide practical benefits and ultimately lead to more robust solutions. 
As an example, in medicine, a reanalysis of data on hip fractures among the elderly~\cite{Caillet:2015hip} found that 
the causal analysis was able to identify which factors mediate the effect of the others and to what extent, in addition to providing predictions on par with traditional methods.
In another study, Richens et al.~\cite{Richens:2020improving} showed that medical diagnosis based on causal inference performed almost twice as good (25th percentile vs. 48th of the performance of human doctors) as the classical, associative\slash statistical method.

A key element of causal inference is to use directed acyclic graphs (DAGs) to model the observed factors' dependence structure. Nodes in the graph denote factors, and edges denote ways that (parent) variables cause changes in other (child) factors. The DAG of a causal model thus makes the causal dependence structure explicit. 
In addition to the DAG, a causal model includes a way to estimate 
child factors based on the values of their parent(s).
In a so called structural equation model, this is achieved via equations, typically linear but more general forms can be used, that relate each child factor to the nodes from which is has incoming DAG edges. 
Another model variant, the probabilistic causal model, uses a single probability distribution over the factors~\cite{Pearl:2009aa}.
While the causal structure, i.e., the DAG, is sometimes known or can be formulated from a theory, it is more common to use so called structural learning (also called causal discovery) to identify the causal structure from data~\cite{Spirtes:2016causal}.
Structural learning methods can either work only from observational data or also (or solely) use interventional data, i.e. when we have taken actions that lead to changed values for some subset of factors of a DAG.
A hallmark of causal analysis is that its DAGs can be used both to guide which factors to intervene on, i.e. change, 
and then tell us how to calculate the causal effects from our observations.

\subsection{Overview of CPDA}

Causal Program Dependence Analysis (CPDA) aims to model and quantify the strength
of program dependence relations.
The dependence reported by CPDA is thus not 
binary: rather, it represents how likely a change to the value of a program element $S_i$ is to 
\emph{cause} a change to the value of another element $S_j$. 
To capture the causal relations 
between program elements, we initially capture the behavior of the original 
program when executed on a test suite, to use as oracle. 
We then observe whether systematically mutated program variants behave 
differently to the oracle. Technically, we let the change 
status of a program element be a random variable in causal inference.
Our causal inference thus uses both purely observational data, 
the behavior of the unmutated program, as well as interventional data,
the behaviors of the mutated program variants.

Figure~\ref{fig:framework} shows the overall framework of CPDA. Given a 
program, CPDA identifies target program elements to analyze, and 
subsequently mutates and instruments the code (Sec.~\ref{sec:instrumentation}). The 
instrumentation allows us to generate an \emph{oracle}, a set of recorded 
behaviors, i.e. states of variables, of the original, unmutated program. 
Super-mutation allows us to apply interventions to the program elements and 
monitor the resulting changes~\cite{Untch:1993ab}. 
Using these observations (Sec.~\ref{sec:change_representaiton}), CPDA builds a \emph{causal structure} (Sec.~\ref{sec:causal_discovery}) 
and performs causal inference to calculate the causal and the direct dependence 
(Sec.~\ref{sec:cpdm}). 
We also introduce Causal Dependence based Fault Localization (CDFL), a novel 
fault localization technique based on CPDA. Using causal dependence, we try to 
identify the program element that is the most likely to cause the observed 
failure (Sec.~\ref{sec:cdfl}).

\subsection{Instrumentation, Oracle, and Mutation}
\label{sec:instrumentation}

We mutate and instrument the target program simultaneously, so that we can 
apply intervention to the behavior of the original program and observe its 
impact. Given a program, let a \emph{node} correspond to each left-hand side 
variable that occurrs in assignment statements, function parameters, 
predicates, and return values of functions: we use the term ``node'' because CDPA 
eventually will represent all these program elements as nodes in the graph of a
causal model. 
Given a node, its \emph{trajectory} represents 
the behavior of the node, i.e. the sequence of values it takes during execution.

\subsubsection{Instrumentation}

A pair consisting of an intervention and the measurement of its impact is called an 
observation. To efficiently produce a large and diverse set of observations 
for program $P$, we construct a \emph{super mutant}~\cite{Untch:1993ab}, i.e., 
a meta-mutated program that takes a mutation position (a unique index of each 
node), and a mutation value that will be used whenever the node is executed, 
as input: this reduces the number of compilations required for multiple 
mutations. 
The instrumentation begins by indexing all 
the nodes in the target program using a parser, and injecting a helper 
function. The helper function, when given a target node index, will overwrite 
(i.e., mutate) the value of the node, and log the result of mutation, during 
execution. For all the other nodes, the helper function simply logs their 
current value observed during execution (i.e., record their trajectories).

\subsubsection{Oracle}

Once the target program is instrumented, we can obtain the oracle for each 
node using the given test suite. An oracle for a node is simply a collection of 
all of its trajectories, one per a test input in the test suite, when no mutations are active during execution.

\subsubsection{Mutation}

CPDA currently targets primitive variables of the following types: bool, char, 
int, long, float, double, and string\footnote{Some ambiguous types in the 
subject programs have been resolved manually}. For booleans, we mutate by simply 
negating the 
original value. For primitive types, we sample a random value from a Gaussian 
distribution that is based on all observed values in its trajectories. 
However, when the domain knowledge clearly specifies the range of possible 
values (such as an input parameter that can be only 0, 1, or 2), we simply 
sample from the given range with a uniform probability.
Finally, for strings, we first sample the string length from a Gaussian 
distribution that is based on the length of all observed strings for the node, and 
subsequently sample a random lowercase string of that length. We sample the 
value mutation up to $N_{mpn}$ times per node.\footnote{Due to boolean nodes or nodes with predetermined value ranges, we may not be able to sample the same number of mutated values for all nodes.}
We leave more refined data mutation and generation strategies for future work
and note that techniques for test generation can likely be of use as well as 
able to handle more complex, structured data types~\cite{Feldt:2013finding}.

\subsection{Representing Behavioral Changes}
\label{sec:change_representaiton}

Once we collect the oracle (trajectories obtained by executing the original 
program with the given test suite) and our observations (trajectories obtained 
by executing the mutated programs with the given test suite), we can identify 
the change of program behavior due to value mutations. Formally, we define
\emph{change of the behavior} for a node as follows:

\begin{definition}[Change of the behavior of a node]\label{def:bhv_chg}
Behavior of node $S_i$ in program $P$ has changed for a given mutation and an 
input, if the trajectory for $S_i$ using the mutated program $P'$ is different 
from the trajectory of $S_i$ using $P$.
\end{definition}

From now on, let us overload the notation $S_i$ to represent both the node 
itself and the variable denoting whether the behavior of $S_i$ has changed under 
$P'$ or not; $S_i = 1$ if changed, 0 otherwise. For a fixed mutated version $P'$ 
and an input, we get a single \emph{observation}, which is a boolean vector 
whose length is equal to the number of nodes in the program: each value in the 
observation indicates whether the behavior of the corresponding node has 
changed or not. Given a set of observations, we can calculate the probability 
of a node changing its behavior.

\begin{definition}[Probability of the behavior change]\label{def:prob_bhv_chg}
  Given Node $S_i$ from Program $P$ and a set of observations $O$, $P_O(S_i = 1)$ represents the probability of a change in $S_i$'s behavior, given $O$, is:
\[
    P_O(S_i = 1) = 1 - P_O(S_i = 0) = \frac{\left| \left\{ \mathit{obs} \in O \mid \mathit{obs}[S_i] = 1 \right\} \right|}{\left| O \right|}\,.
\]
\end{definition}

Moreover, we can also express the conditional probability of node $S_j$ changing its behavior given that node $S_i$ has changed its behavior:

\begin{definition}[Conditional probability of the behavior change]\label{def:cond_prob_bhv_chg} For a given set of observations, $O$, behavior, and 
$O^* = \{o \mid o \in O \land$ node $S_j$ is not mutated in the mutated program that generates $o\}$:
\[
\begin{split}
P_O(S_j = 1 \mid S_i = 1) &= \frac{P_O(S_j = 1 \land S_i = 1)}{P_O(S_i = 1)} \\
        &= \frac{\left| \left\{ \mathit{obs} \in O^* \mid \mathit{obs}[S_j] = 1 \land \mathit{obs}[S_i] = 1 \right\} \right|}{\left| \left\{ \mathit{obs} \in O^* \mid \mathit{obs}[S_i] = 1 \right\} \right|}\,.
      \end{split}\]

\end{definition}

The conditional probability in Definition~\ref{def:cond_prob_bhv_chg} becomes 
the foundation of causal dependences that we calculate later. We exclude the 
mutation on $S_j$ itself, hence $O^*$, when calculating the conditional 
probability $P_O(S_j = 1 \mid S_i = 1)$ . This is to avoid negating the effect 
from the change of $S_i$ with a mutation of $S_j$ itself.

Each node can have up to $N_{mpn}$ mutated values (see 
Section~\ref{sec:instrumentation}). Since each node may have a different 
number of mutated 
values, we normalize the probabilities in Def.~\ref{def:prob_bhv_chg} 
and~\ref{def:cond_prob_bhv_chg} by multiplying the reciprocal of the number of 
sampled mutated values. For example, each observation for an integer type node 
mutation from ten mutation values has a weight of 0.1. 

\subsection{Causal Structure Discovery}
\label{sec:causal_discovery}

The conditional probability defined in Def.~\ref{def:cond_prob_bhv_chg} 
expresses the \emph{association} between changes of program elements: changes 
are simply observed together. To elevate this to \emph{causal} inference, we 
need the concept of a specific change \emph{preceding} another. It is also 
necessary to distinguish between direct predecessors, i.e., nodes whose change 
directly affects our target node, and non-direct predecessors, i.e., 
nodes whose change does affect our target node but indirectly through one or 
more direct predecessors. 

A \emph{causal structure} allows us to introduce this concept of precedence. Program 
dependence is inherently a form of causal precedence: if node $S_j$ depends on 
node $S_i$, a change in $S_i$ will precede the change in $S_j$. Consequently, 
in principle, a perfectly accurate PDG of the target program can serve as a 
causal structure, if available. However, in reality, existing dependence 
analysis is both computationally costly and may not be accurate. Instead, we 
use ideas from the causal inference field and approximate the causal structure 
from the set of observations we have collected.

Let us first formally define the predecessors of a node. We identify the 
predecessors of a node $S_j$ by checking whether intervening on a particular 
node affects $S_j$. We call predecessors of a node as its \emph{
intervention parents}.

\begin{definition}[Intervention Parent]
For a program $P$ and a set of nodes $S$ from the program, a set of nodes 
$\mathit{IPA}_j \subset S$ is said to be intervention parents of $S_j$ if 
mutating any node in $\mathit{IPA}_j$ changes the behavior of $S_j$ for at 
least one input. In other words, if a mutated program $P'$ that changes the 
value of $S_k$ to $v$ causes $S_j = 1$ under input $i$, then $S_k\in\mathit{IPA}_j$.
\end{definition}

However, the precedence relationship may be direct, or indirect (i.e., a 
change in one of the intervention parents will, by definition, precede the 
change in the child node, but its effect may or may not need to pass through another 
parent to affect the child node).
Based on the theory of causal inference, we define \emph{Markovian parents} of a node as the 
minimal set of direct predecessors among the intervention parents. 

\begin{definition}[Markovian parent]\label{def:markovian}

The Markovian parents of $S_j$, $\mathit{PA}_j$, is a minimal set
predecessors of $S_j$ that renders $S_j$ independent of all its intervention 
parents. In other words, $\mathit{PA}_j$ is any subset of $\mathit{IPA}_j$ such 
that it satisfies $P(s_j \mid \mathit{pa}_j) = P (s_j 
\mid \mathit{ipa}_j)$ while no other proper subset of $\mathit{PA}_j$ 
does.\footnote{\label{foot:realization}Lowercase symbols (e.g., $x, \mathit{pa}_j$, and $\mathit{ipa}_j$) 
denote particular realizations of the corresponding variables (e.g., 
$X, \mathit{PA}_j$, and $\mathit{IPA}_j$).}
\end{definition}

Subsequently, the Markovian parent-child relation constructs the causal 
structure of the program.

\begin{definition}[Causal structure]
The causal structure of a program represents the Markovian relation between 
the nodes. The causal structure is represented by a graph, $G = (V, E)$, where 
$V = \left\{S_i\right\}_{i\in{1..n}}$ and $E = \left\{\left(S_i, S_j\right) 
\mid S_i \in \mathit{PA}_j \right\}$, where $\mathit{PA}_j$ denotes Markovian 
parents of $S_j$.
\end{definition}

\begin{algorithm}[ht]
  \small
  \DontPrintSemicolon
  \SetKw{Break}{break}
  \KwIn{$\mathit{IPA}_j$: Intervention parents of $S_j$, \newline
        $Dist$: Ordered nodes considering the distance from $S_j$, \newline
        $O$: Observations generated from inputs that cover $S_j$ in the original program.}
  \KwOut{$\mathit{PA}_j$: Markovian parents of $S_j$}
  $\mathit{PA}_j \leftarrow \{\}$ \\
  $Cand \leftarrow \mathit{IPA}_j$ \\
  \While{$Cand.len() \neq 0 \land Cand \neq \mathit{PA}_j$}{
    $Remain \leftarrow Cand \setminus \mathit{PA}_j$ \\
    $S_d \leftarrow Dist.getFarthest(Remain)$ \\
    \eIf{$Cand.len() == 1$}{
      \If{$P_O(S_j = 1 \mid S_d = 0) = P_O(S_j = 1 \mid S_d = 1)$}{
        $is\_parent \leftarrow False$
      }
      \lElse{
        $is\_parent \leftarrow True$
      }
    }{
      $S_{other} \leftarrow Cand \setminus S_d$ \\
      $is\_parent \leftarrow False$ \\
      \ForEach{$s_{other} \in O|_{S_{other}}$}{
        \If{$P_O(S_j = 1 \mid S_d = 0, s_{other}) \newline \neq P_O(S_j = 1 \mid S_d = 1, s_{other})$}{
          $is\_parent \leftarrow True$ \\
          \Break
        }
      }
    }
    \lIf{$is\_parent$}{
      $\mathit{PA}_j.add(S_d)$
    }
    \Else{
      $Cand.remove(S_d)$ \\
      $\mathit{PA}_j \leftarrow \{\}$
    }
  }
  \Return{$\mathit{PA}_j$}
  \caption{Generate Markovian parents from $\mathit{IPA}$}
  \label{alg:mrkv}
\end{algorithm}

It requires an exponential number of computations to compute the conditional probability in Def.~\ref{def:markovian} for every possible combination of intervention parents. Thus, we approximate the Markovian parents by iteratively removing non-Markovian parents from intervention parents.
Algorithm~\ref{alg:mrkv} shows the process of removing non-Markovian parents 
from $\mathit{IPA}_j$.
We choose one node $S_d$ from $\mathit{IPA}_j$ and check whether $S_j$ is 
independent from $S_d$, given all other candidate nodes. 

If $S_d$ is the only candidate node left (Line 7-9), we check whether $S_j$ is 
independent from $S_d$. If it is, $S_d$ is not a Markovian parent, otherwise, 
$S_d$ is not a Markovian parent. If there are other candidate nodes $S_{other}$ 
(Line 11-16), we check the conditional independence of $S_j$ from $S_d$ for every 
realization of $S_{other}$. If $S_j$ is conditionally independent from $S_d$ 
for all $s_{other}$, $S_d$ is not a Markovian parent. Otherwise, $S_d$ 
\emph{could be} a Markovian parent of $S_j$. This is because $S_j$ can be 
conditionally independent from $S_d$ if $S_{other}$ changes. Therefore, we 
re-check all possible Markovian parents every time we find a new non-Markovian 
parent. To minimize re-checking cost, we order the candidate nodes ($Dist$) to 
choose the most unlikely to be the Markovian parent first (Line 5). Our 
intuition is that, during the execution of the program, the node executed 
earlier than the execution of $S_j$ is less likely to be a parent of $S_j$. 
Algorithm~\ref{alg:mrkv} \emph{approximates} Markovian parents of a node, as 
it only considers individual nodes, and not their combination, for Makovian 
parents.

The last step of building a causal structure is to eliminate cycles in the 
structure. In causal inference, the form of the causal structure is the 
\emph{Bayesian network}, which has no cycles. To eliminate cycles, we remove 
from $S_i$ in $\mathit{PA}_j$ if $S_j$ is in $\mathit{IPA}_i$.\footnote{We 
further discuss the cyclic relation in the causal model in 
Section~\ref{sec:discussion}.}

\subsection{Causal Program Dependence Model}
\label{sec:cpdm}

We first introduce causal dependence, which measures the total effect of each 
node's change that causes a change in another node, and subsequently introduce 
direct dependence, which measures the effect of one node on another excluding 
all the (indirect) effects through other nodes.

\subsubsection{Causal dependence}

The conditional probability $P(\cdot | x)$\footnoteref{foot:realization} represents the probability 
when one \emph{observes} that $X$ has value $x$. By having a causal structure, we can 
instead estimate the probability when we \emph{force} $X$ to have the value $x$, denoted as 
$P(\cdot | do(x))$. The difference between 
"forcing" and "observing" comes from whether and how the effect of $X$ is affected by other nodes. 
A \emph{confounding bias}, a distortion representing the event that is 
associated but not casually related to the observation, appears through the 
so-called \emph{backdoor path} in a causal DAG~\cite{Pearl:2009aa}. By ignoring the incoming 
effect of $X$, we can remove the effect through the backdoor path to $X$, 
subsequently eliminating the confounding bias from the association between $X$ 
and another node $Y$. Note that backdoor paths in CPDM are based on the inferred causal structure, unlike existing work~\cite{Baah:2010aa}
Based on Pearl et al.~\cite{Pearl:2009aa}, the causal effect estimating 
$P(\cdot|do(x))$ is formally defined below: we use a set of nodes, instead of 
a single node, to include the more general scenario where multiple nodes are 
mutated simultaneously. 

\begin{definition}[Causal Effect]
\label{def:causal_effect}
Let $G = (V, E)$ be a causal structure. Given two disjoint sets of nodes, $X,Y 
\subset V$, the causal effect of $X$ on $Y$, denoted either as $P(y \mid 
\hat{x})$ or $P(y \mid do(x))$, is a function from $X$ to the space of 
probability distributions on $Y$. For each realization $x$ of $X$, $P(y \mid 
\hat{x})$ gives the probability that $Y = y$ induced by deletion from the 
causal structure of all the edges to nodes in $X$ other than $x$ and thus 
considering $X = x$. The causal effect $P(y \mid do(x))$ is calculated as:
\begin{equation*}
  P(y \mid do(x)) = \sum_{pa_X} P(y \mid x, pa_X) P(pa_X)\,,
\end{equation*}
where $PA_X$ represents the Markovian parents of $X$.
\end{definition}

In causal program dependency analysis, we aim to measure the amount of impact 
that the change of one node $S_i$ has on another node $S_j$, i.e., the 
difference between the causal effect from ``$S_i = 1$'' to $S_j$, and 
from ``$S_i = 0$'' to $S_j$. Thus, we introduce \emph{causal dependence} as follows:

\begin{definition}[Causal Dependence]
The causal dependence of node $S_i$ to $S_j$ measures how much changing the 
behavior of $S_i$ affects the behavior of node $S_j$. Given a set of 
observations $O$, the causal dependence of $S_i$ to $S_j$, $\mathit{CD}_O(S_i, S_j)$, is defined as:
\begin{equation*}
    \mathit{CD}_O(S_i, S_j) = P_O(S_j = 1 \mid do(S_i = 1)) - P_O(S_j = 1 \mid do(S_i = 0))\,.
\end{equation*}
\end{definition}

\subsubsection{Direct dependence}

\emph{Direct dependence} quantifies the amount of effect that is not mediated by, 
any other nodes. More formally, it measures the sensitivity of $Y$ 
to changes in $X \in \{$Markovian parents of $Y\}$ while all 
other Markovian parents of $Y$ are held fixed. 
Based on Pearl et al.~\cite{Pearl:2009aa}, a generic definition that fits such sensitivity is the natural direct effect.

\begin{definition}[Natural Direct Effect]
The natural direct effect, denoted either as $\mathit{NDE}_{X: x \rightarrow x'}(Y)$ is the expected change in $Y$ induced by changing $X$ from $x$ to $x'$ while keeping all mediating factors constant at whatever value they would have been under $do(x)$. The natural direct effect, $\mathit{NDE}_{X: x \rightarrow x'}(Y)$, is calculated as:
\begin{equation*}
   \sum_{z} \left[E\left(Y \mid do(x', z)\right) - E \left(Y \mid do(x, z)\right)\right] P\left(z \mid do(x)\right),
\end{equation*}
where $Z$ represents all parents of $Y$ excluding $X$.
\end{definition}

Similar to causal dependence,
in the context of CPDA, we estimate how much effect a parent has on a 
child node $S_j$ when the parent node $S_i$ alter from ``unchanged'' to 
``changed'', regardless of all other parents of $S_j$.
Based on the definition of the natural direct effect, we define the direct dependence as follows.

\begin{definition}[Direct Dependence]
  \label{def:DD_def}
  The direct dependence of $S_j$ from $S_i$, denoted as $\mathit{DD}_I(S_i, S_j)$, is the average of the natural direct effect of $S_i$ towards $S_j$ over all inputs $I$:
  \begin{equation*}
    \mathit{DD}_I(S_i, S_j) = \frac{1}{|I|}\sum_{t \in I} \mathit{NDE}_{O_t, S_i: 0 \rightarrow 1}(S_j)\,,
  \end{equation*}
where $O_t$ denotes the observations from input $t$, and $\mathit{NDE}_{O_t, S_i: 0 \rightarrow 1}(S_j)$ denotes the natural direct effect using the observations $O_t$.
\end{definition}

\subsubsection{Causal Program Dependence Model}

Given the definition of the direct dependence, CPDM is a weighted dependence graph 
of the program. Its structure is the causal structure, and the edge's weight, 
indicating the strength of dependence, is the direct dependence between the nodes. 
CPDM is thus a novel graph representation to explain program dependence in a 
continuous, gradual way.

\subsection{Causal Dependency Based Fault Localization}
\label{sec:cdfl}

CPDA allow us to compare the relative strengths of different dependency 
relationships in a program. To highlight the benefits of this quantification, 
we propose Casual Dependence based Fault Localization 
(CDFL). The main hypothesis of CDFL is that the faulty program element, that we are trying to locate, 
will have more impact on the output element in failing executions than in passing executions.
This is intuitive since the basic assumption in fault localization is that there is a faulty
element that is causing the faulty output; if it is faulty but is not causing this particular fault
we are (currently) not looking for it.
The ability to compare the strength of different dependence 
relationships means that we can \emph{rank} program elements according to their effect on 
the faulty output. This is not possible in fault localisation techniques 
based on binary program dependence, such as dicing~\cite{Agrawal:1990aa}.

Given an output node $S_{out}$ which causes the test failure, we can compute 
the suspiciousness score of node $S_i$, $susp(S_i)$, as:

\begin{equation*}
   \frac{1}{|I_{fail}|} \sum_{t \in I_{fail}} \mathit{CD}_{O_t}(S_i, S_{out}) - \frac{1}{|I_{pass}|} \sum_{t \in I_{pass}} \mathit{CD}_{O_t}(S_i, S_{out}),
\end{equation*}

\noindent where $I_{fail}$ and $I_{pass}$ denote a set of failing and passing 
test inputs, respectively. Note that the value of $S_{out}$ represents the 
change of outcome, and not pass or fail.

\begin{table}[ht]
\caption{Example comparing CDFL to SBFL}
\label{tbl:cdfl-adv-sbfl}
\scalebox{0.8}{
\begin{tabular}{l|rr}
\toprule
Code                    & \multicolumn{1}{l}{Rank$_\text{SBFL}$} & \multicolumn{1}{l}{Rank$_\text{CDFL}$} \\ \midrule
\texttt{a = 3}          & 1                              & 2   \\
\texttt{b = 4}          & 1                              & 3   \\
\texttt{c = a \% 3 \color{red}+ 1}& 1                    & 1   \\
\texttt{return c}       & -                              & -   \\
\bottomrule
\end{tabular}}
\end{table}

\subsubsection{Advantages over SBFL}

Table~\ref{tbl:cdfl-adv-sbfl} contains a motivating example showing the advantages 
of CDFL over SBFL: the fault is typeset in red. SBFL assigns suspiciousness 
scores to program elements 
based on the coverage and outcomes of test executions~\cite{Wong:2016aa}. 
Consequently, its performance depends significantly on the differences in 
control flow between passing and failing executions: it fails to distinguish 
the failing execution in Table~\ref{tbl:cdfl-adv-sbfl}, which can be only 
characterized in data flow. Moreover, all statements in the same program block 
are assigned with the same suspiciousness score, as can be seen in the 
example. CDFL, however, can correctly analyze that the faulty return value is caused by the assignment to variable \texttt{c}.

\begin{table}[ht]
\caption{Example comparing CDFL to Dynamic Slicing and Dicing. 
`\textvisiblespace' denotes the whitespace.}
\label{tbl:cdfl-adv-dep}
\resizebox{\linewidth}{!}{
\begin{tabular}{lrrr|rr|r}
\toprule
\multirow{2}{*}{Code}                                         & \multicolumn{3}{c|}{Coverage} & \multirow{2}{*}{DS} &     \multirow{2}{*}{Dice}       &  \multirow{2}{*}{Susp$_\text{CDFL}$}            \\
                                                              & ``a'' & ``a\textvisiblespace'' & ``\textvisiblespace a\textvisiblespace'' &   &   &                 \\ \midrule
\texttt{s = input()}                                          & 1     & 1      & 1       & 1 & 0 & 1.0 - 1.0 = 0.0 \\
\texttt{pred = isEndSpace(s)}                                 & 0     & 1      & 1       & 1 & 0 & 1.0 - 0.5 = 0.5 \\
\texttt{if (pred) p = \sout{p.rstrip()} \color{red}p.strip()} & 0     & 1      & 1       & 1 & 0 & 1.0 - 0.5 = 0.5 \\
\texttt{return p}                                             & -     & -      & -       & - & - & -               \\ \midrule
Test Results                                                  & P     & P      & F       &   &   &                 \\
\bottomrule
\end{tabular}
}
\end{table}

\subsubsection{Advantages over Dynamic Slicing and Dicing}

Table~\ref{tbl:cdfl-adv-dep} contains a motivating example showing the 
advantages of CDFL over DS, a dynamic backward slice of the returned variable 
\texttt{p}, and a dice~\cite{Agrawal:1995aa}, which essentially returns the 
set difference between dynamic slices computed using failing test inputs and 
passing test inputs. Intuitively, if there is a dependence relationship that 
is only exercised in failing executions, dicing reports it as the likely root 
cause of the failure. The original code in Table~\ref{tbl:cdfl-adv-dep} 
intends to only remove trailing whitespaces (denoted as `\textvisiblespace'), but the faulty version strips all 
whitespaces. Because one of the passing inputs, \mbox{``a\textvisiblespace''},
and the failing input, \mbox{``\textvisiblespace a\textvisiblespace''}, execute all lines of the code, 
dynamic slicing reports all lines to be equally faulty, while dicing concludes 
that none of them are faulty. However, CDFL reports that the faulty line is 
more suspicious, based on the quantitative dependence.

\section{Experimental Setup}
\label{sec:experimental_setup}

This section presents our research questions and the set-up of our empirical 
evaluation. 

\subsection{Research Questions}
\label{sec:rqs}

CDFL is but one possible use of the general program dependence analysis method of CPDA.
In addition to evaluating CDFL, we thus first ask a more general research question about CPDA.

\noindent\textbf{RQ1. Effectiveness of CPDA:} How well does CPDA capture the 
relative strengths of dependence between program elements? We expect the 
direct dependence to capture the varying degrees of dependence between program 
elements. To investigate this, we build CPDMs for an example program with well 
known dependencies, and examine how it differs from PDG. We also investigate its sensitivity when using different test suites.

\noindent\textbf{RQ2. Performance of CDFL:} How well does CDFL perform 
compared to other, widely studied fault localisation techniques? We 
evaluate the utility of CPDA through CDFL, by comparing the fault 
localisation results to Spectrum Based Fault Localization (SBFL), dynamic 
slicing, and dicing. We also investigate the impact that the number 
of observations, $N_{mpn}$, has on the accuracy of CDFL.

\subsection{Subjects and Baselines}
This section describes the subject programs, and baseline techniques, for the 
CPDA (RQ1) and CDFL (RQ2) study, respectively.

\subsubsection{CPDA Study (RQ1)}

\begin{figure}[ht]
\noindent
\begin{minipage}[t]{.51\columnwidth}
\begin{lstlisting}[numbers=left,numbersep=3.5pt]
def main() {
   <1>characters = 0
   <2>lines = 0
   <3>words = 0
   <4>inword = 0
   <5>_pred1 = getChar(<6>c)
   while (_pred1) {
      <7>characters = characters + 1
      <8>_pred2 = c == '\n'
      if (_pred2)
         <9>lines = lines + 1
      <10>_pred3 = isLetter(c)
      if (_pred3) {
         <11>_pred4 = inword == 0
         if (_pred4) {
            <12>words = words + 1
         }
\end{lstlisting}
\end{minipage}%
\begin{minipage}[t]{.49\columnwidth}
\begin{lstlisting}[numbers=left,firstnumber=18,numbersep=3.5pt]
         <13>inword = 1
      }
      else
         <14>inword = 0
      <15>_pred1 = getChar(<16>c)
   }
}
def isLetter(<17>c) {
   <18>_pred5 = ((c >= 'A' && c <= 'Z') 
      || (c >= 'a' && c <= 'z'))
   if (_pred5)
      <19>_ret = True
   else
      <20>_ret = False
   return _ret
}
\end{lstlisting}
\end{minipage}
\caption{Pseudo-code of the word count program}
\Description{Pseudo-code of the word count program}
\label{fig:wc}
\end{figure}

To investigate RQ1, we apply CPDA to the word count (\wc) program, 
which has been widely studied for program dependence analysis~\cite{
Gallagher:1991aa,Lee:2019aa}. Figure~\ref{fig:wc} shows the pseudo-code of the 
word count program. The node index annotations in angled brackets are used 
to refer to corresponding nodes in Section~\ref{sec:rq1}.
The word count program takes a text as input and counts its number of characters, lines, and words.
We use a test suite of 15 test cases, which can be grouped into the following:

\begin{description}
\item [\emph{onechar}]: Tests 1-3 contain a single letter.
\item [\emph{oneword}]: Tests 4-7 contain a single word of multiple letters.
\item [\emph{oneline}]: Tests 8-10 contain a single line with multiple words.
\item [\emph{multiline}]: Tests 11-15 contain multiple lines of multiple words.
\end{description}
For each node, we use a sample of 100 mutations ($N_{mpn} = 100$) and observe 
the changes at other nodes.

\begin{table}[ht]
\centering
\caption{Statistics of the Siemens suite}
\label{tbl:siemens}
\scalebox{0.8}{
\begin{tabular}{lrrrrr}
\toprule
Subject & SLoC & N$_\text{bug}$ & N$_\text{test}$ & N$_\text{failing test}$ & N$_\text{node}$ \\ 
\midrule
\tcas & 135 & 34 & 5--7   & 1--1 & 43  \\
\sch  & 292 & 4  & 8--9   & 1--2 & 94  \\
\scht & 297 & 3  & 7--10  & 1--1 & 147 \\
\tot  & 346 & 18 & 7--10  & 1--4 & 172 \\
\prt  & 475 & 3  & 11--17 & 1--1 & 293 \\
\prtt & 401 & 6  & 12--15 & 1--3 & 186 \\
\rep  & 512 & 24 & 16--20 & 1--2 & 318 \\ \midrule
total & 2,458    & 92 &   -  &  -   & 1,253 \\
\bottomrule 
\end{tabular}}
\end{table}

\begin{figure*}[ht]
    \centering
    \subfigure[PDG]{\includegraphics[width=.24\textwidth,trim={40 40 40 40},clip]{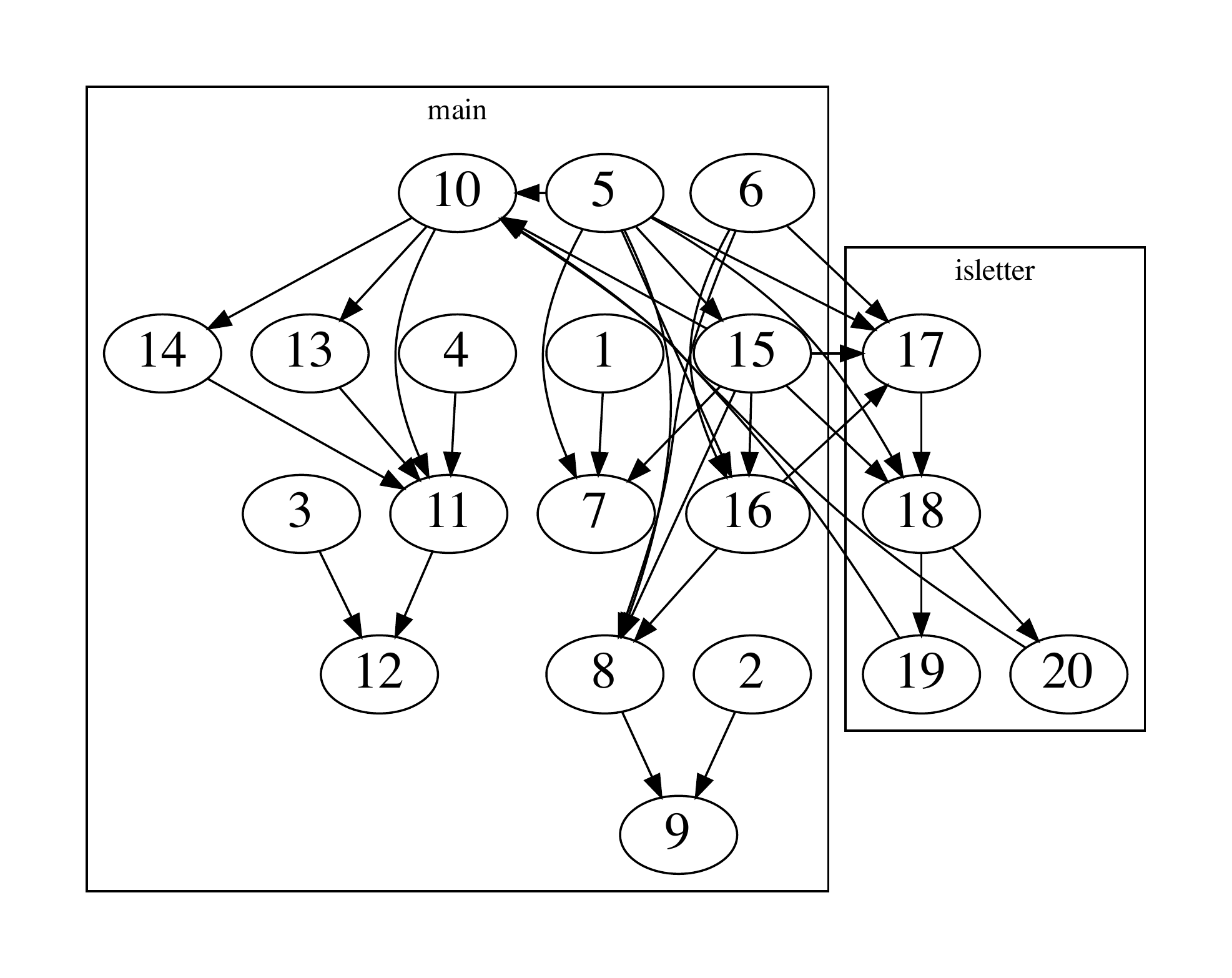}\label{fig:pdg}}\hfill
    \subfigure[$\mathit{DD}>0.2$]{\includegraphics[width=.24\textwidth,trim={40 40 40 40},clip]{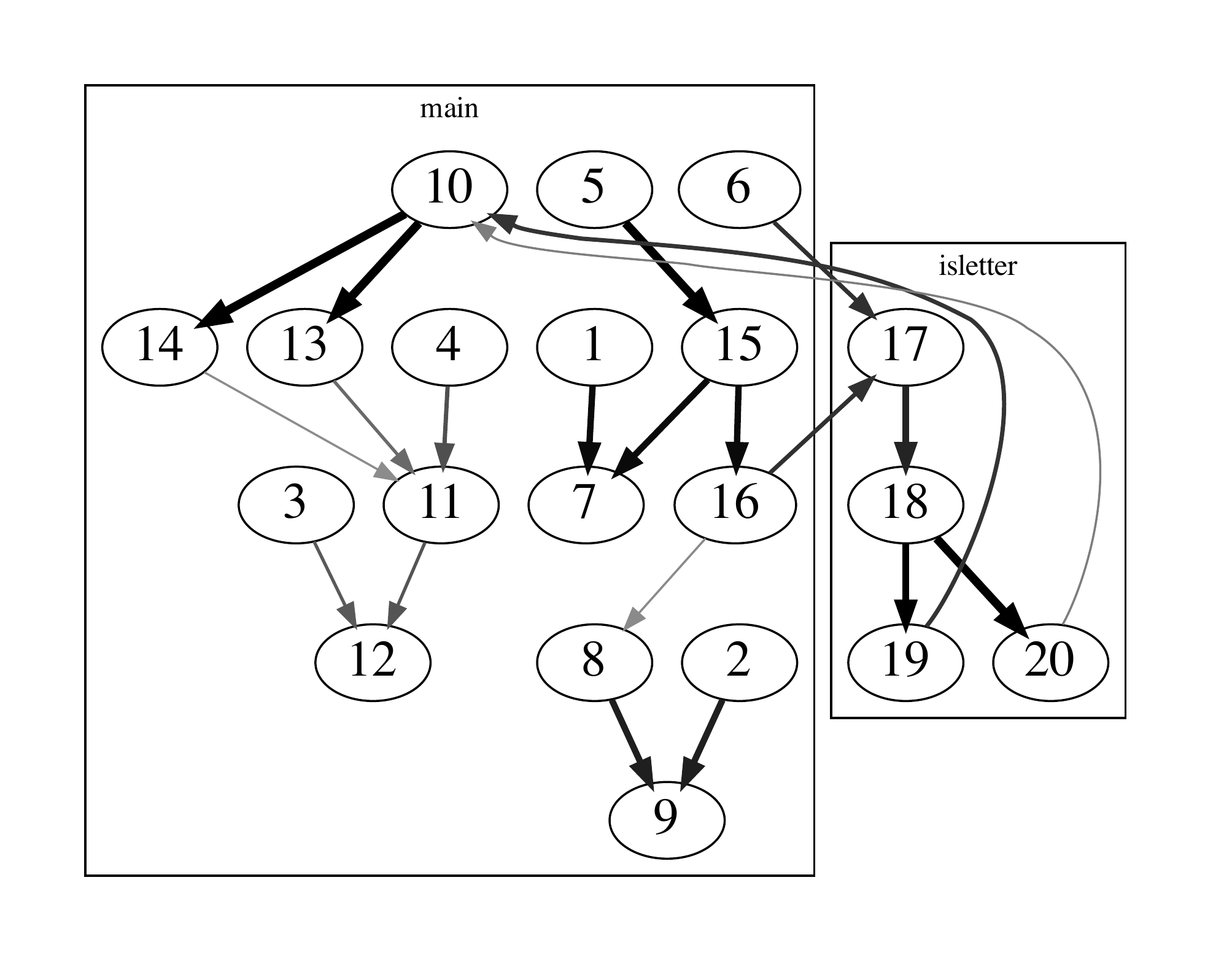}\label{fig:cpdm-2}}\hfill
    \subfigure[$\mathit{DD}>0.8$]{\includegraphics[width=.24\textwidth,trim={40 40 40 40},clip]{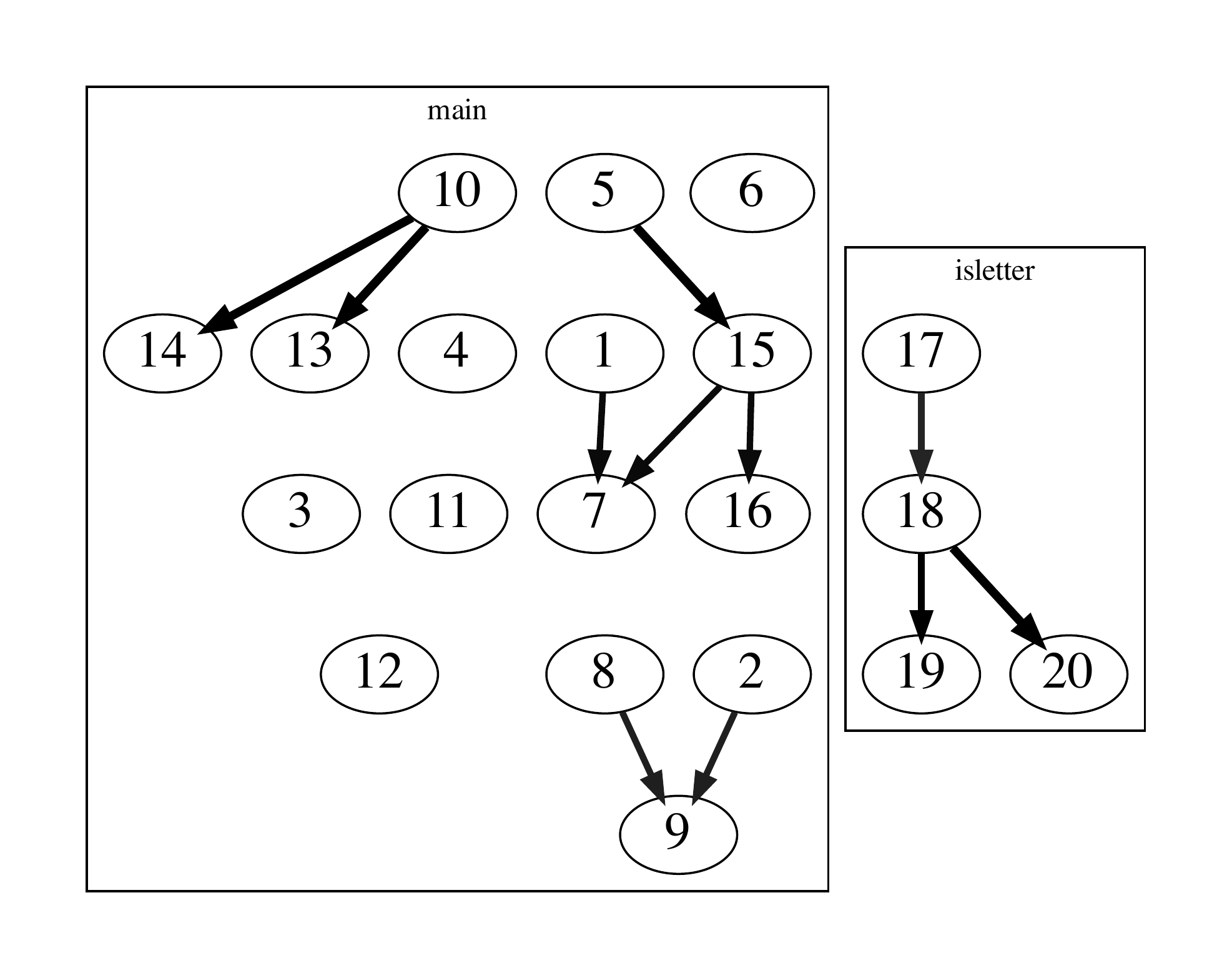}\label{fig:cpdm-8}}\hfill
    \subfigure[$0.8>\mathit{DD}>0.2$]{\includegraphics[width=.24\textwidth,trim={40 40 40 40},clip]{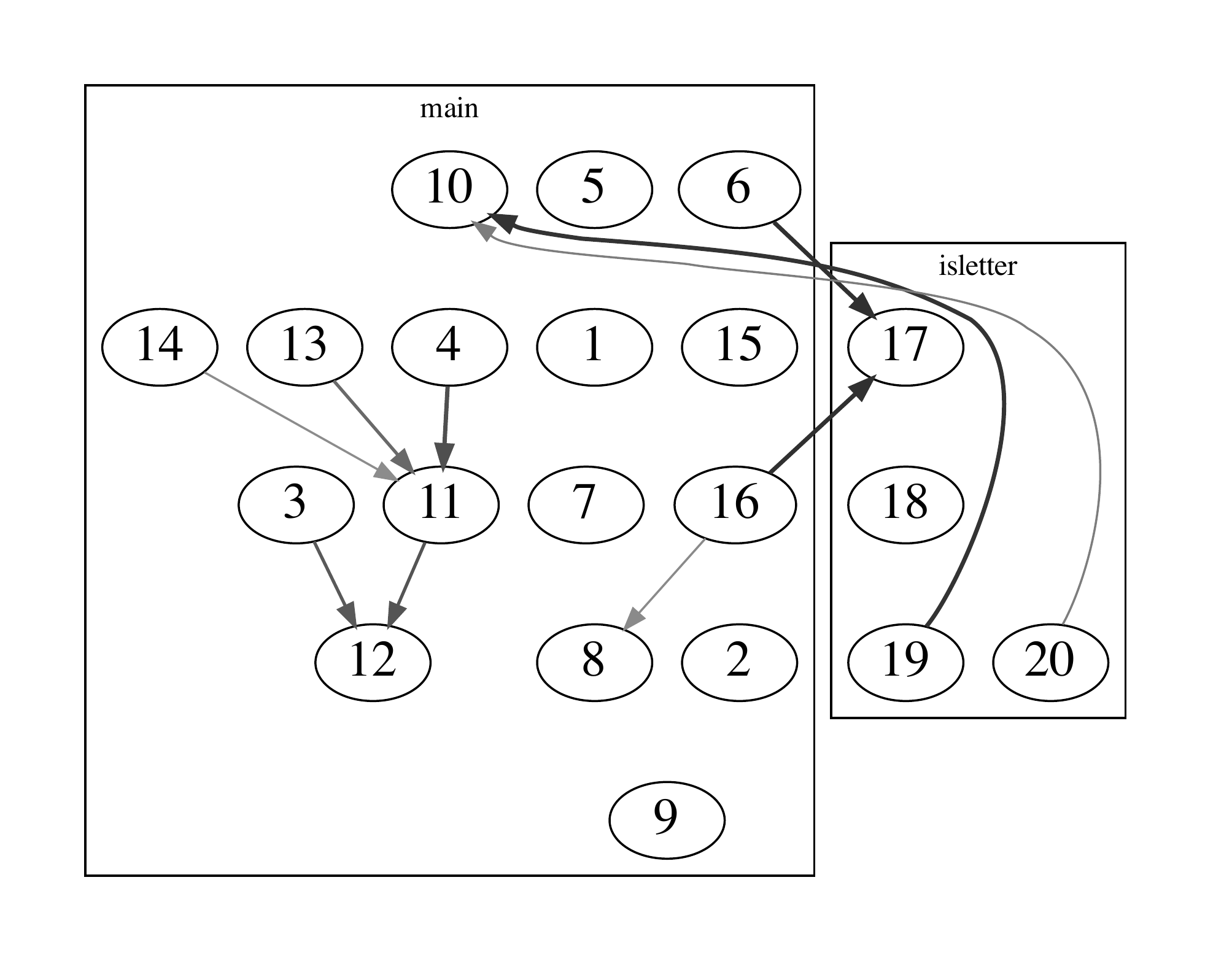}\label{fig:cpdm-28}}
    \caption{Program Dependence Graph (PDG) and Causal Program Dependence Models of word count with different thresholds for the Direct Dependences (DD) For CPDMs, the thicker and darker the edge is, the larger the direct dependence.}
    \Description{Causal Program Dependence Model of word count with different thresholds}
    \label{fig:wc-cpdm}
    \end{figure*}

\subsubsection{CDFL Study (RQ2)}

We compare CDFL to three existing FL techniques: Spectrum Based Fault 
Localization (SBFL), Dynamic Slicing (DS), and Dicing. As SBFL technique, we 
choose Ochiai~\cite{Ochiai:1957fj} which is widely studied in the literature~\cite{Abreu:2007aa}. 

We compute dynamic slices using the dynamic slicing algorithm of
Agrawal and Horgan~\cite{Agrawal:1990aa}, which intersects coverage 
information from a set of inputs with a static slide. We employ \texttt{gcov}
to collect coverage information, and \emph{CodeSurfer}, a static analysis tool 
produced by GrammaTech~\cite{Grammatech-Inc.:2002aa}, for static backward 
slices. Finally, we compute the dice as the difference of the dynamic slices 
of the failing and passing tests.  Note that CDFL ranks nodes, while SBFL ranks program statements. 

We compare four fault localization techniques using programs from the Siemens 
suite~\cite{Do:2005aa}. Table~\ref{tbl:siemens} provides descriptive 
statistics of the studied programs: columns represent Non-comment-non-blank 
Lines of Code (SLoC), number of studied bugs, tests, failing tests, and nodes, 
respectively. We randomly choose a statement coverage adequate test suite 
provided by SIR that contains at least one failing test for each bug. 

Some of the seeded faults provided by SIR have been excluded from our study. 
Excluded faults can be categorized into four groups. First, we exclude 
omission faults, because their root causes do not exist in the given source 
code. Second, we exclude faults for which no failing test case exists. Third, 
we exclude faults that results in nondeterministic behavior, or segmentation 
faults, as they produce either inconsistent, or no coverage information. 
Finally, we exclude any fault whose root cause falls outside the scope of CPDA 
(for example, if the fault is calling a wrong function without storing the 
return value, we cannot represent the fault as a node in CPDA).

For evaluation, we use the widely adopted $acc@n$ metric, which counts the 
number of faults whose root causes are ranked within the top $n$ places. If 
there are multiple faulty elements, we compute $acc@n$ using the most highly 
ranked element. Since SBFL techniques tend to produce a lot of ties, we 
report rankings obtained by multiple tiebreakers.
The average tiebreaker computes the average rank of the tied candidates. Min 
and max tiebreakers rank all tied elements at the highest, and the lowest, 
place, respectively. Finally, line order tiebreaker breaks ties independently 
from score distributions using line numbers, facilitating fair comparison 
between rankings theoretically~\cite{Xie:2013uq}.

\subsection{Implementation \& Environment}

For node selection, we use the open source C static analyzer 
Frama-C~\cite{Kirchner:2015aa}. To insert logging functions in the original 
program, we use srcML~\cite{Collard:2013aa}, an open source tool that parses 
and converts source code into the XML format for manipulation. 

The Markovian parent calculation (Algorithm~\ref{alg:mrkv}) can be costly when 
the number of intervention parents and the number of observations are large. 
To reduce the cost of calculating the Markovian parents, we use \emph{safe 
parents} ($\mathit{SPA}$), which is a subset of the intervention parents.
Given $\mathit{IPA}_j$, the intervention parent set of $S_j$, the safe parent 
set $\mathit{SPA}_j$ is calculated by following heuristics:

\begin{itemize}
\item $\mathit{SPA}_j$ contains all the nodes in the same method as $S_j$ 
($M_j$) or the parameters of the methods calling $M_j$.
\item If $S_j$ is the left-hand side of an assignment statement, and if the 
assignment statement invokes the function $f$, $\mathit{SPA}_j$ contains the 
return nodes of $f$.
\item If $S_j$ uses a global variable, $\mathit{SPA}_j$ contains all the nodes 
of the global variable.
\end{itemize}

While using the safe parent set requires an extra call graph analysis, it 
reduces the Markovian parent calculation time five-fold. Our experiments were 
performed under CentOS Linux 7, on an Intel(R) Xeon(R) CPU E5-2630 v4 with 
250GB of memory.

\section{Results}
\label{sec:results}

This section answers the research questions based on the results of our 
empirical investigation of CPDA. 

\subsection{Effectiveness of CPDM}
\label{sec:rq1}

We first consider the Causal Program Dependence Model created using the entire 
test suite of 15 test cases and then the four models created using the four groups.
Figure~\ref{fig:wc-cpdm} contains the PDG of \wc produced manually
(Fig.~\ref{fig:pdg}), along with visualizations of three CPDMs that contain edges 
with different ranges of direct dependence (DD) values. The thickness of edges 
corresponds to its direct dependence. Figure~\ref{fig:cpdm-2}, ~\ref{fig:cpdm-28},
and ~\ref{fig:cpdm-8}, show edges with direct dependence less than 0.2, between 0.2 and 0.8 (non-inclusive), and greater than 0.8, respectively.

\begin{figure*}[ht]
    \centering
    \subfigure[\emph{oneword} $-$ \emph{onechar}]{\includegraphics[width=.25\textwidth,trim={40 40 40 40},clip]{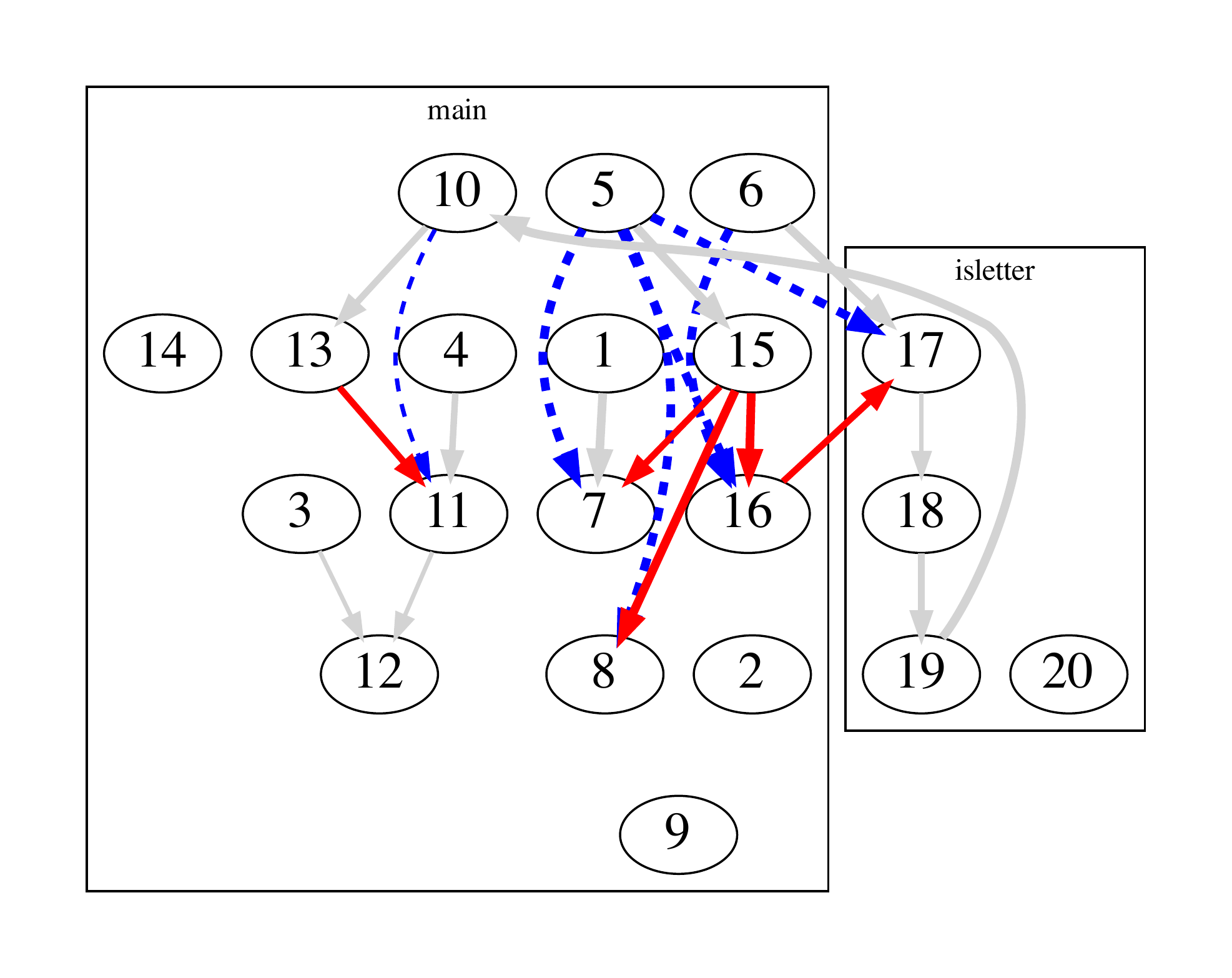}\label{fig:word-char}}\hfill
    \subfigure[\emph{oneline} $-$ \emph{oneword}]{\includegraphics[width=.25\textwidth,trim={40 40 40 40},clip]{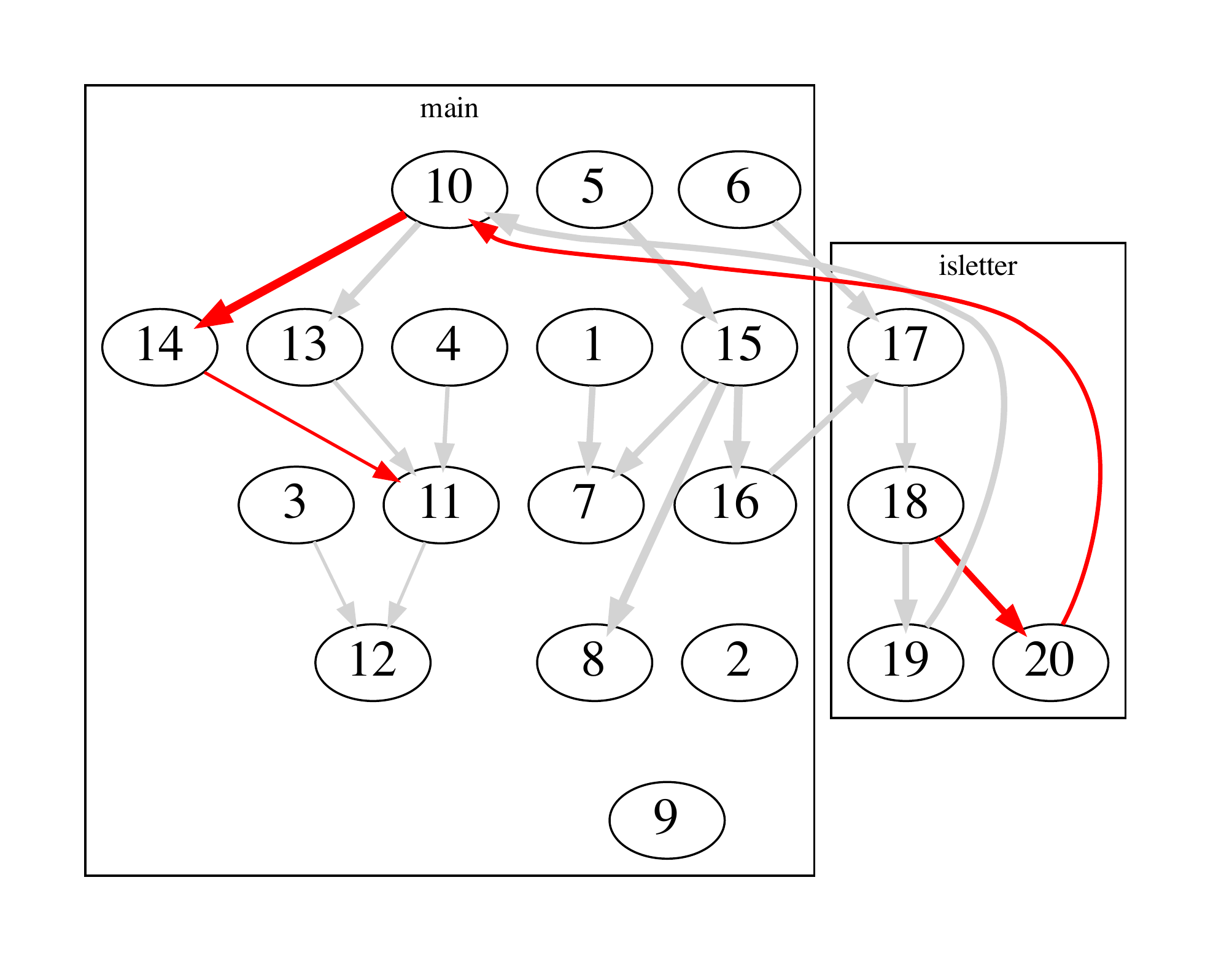}\label{fig:line-word}}\hfill
    \subfigure[\emph{multiline} $-$ \emph{oneline}]{\includegraphics[width=.25\textwidth,trim={40 40 40 40},clip]{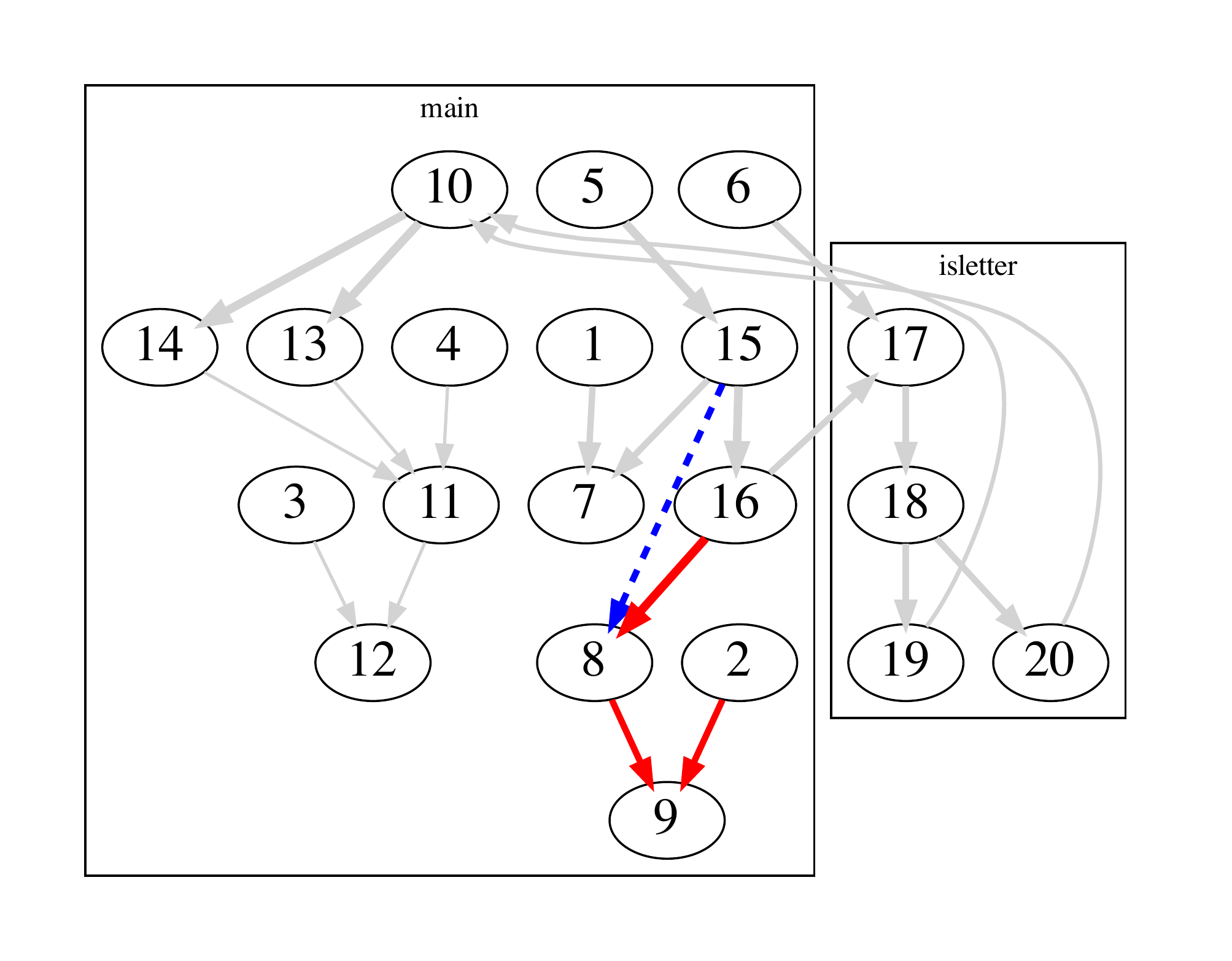}\label{fig:m-1line}}
    \caption{Differences in the CPDMs for word count created using the four different partitions of the test suite.
    In each figure \emph{A} $-$ \emph{B}, shows 
    edges in \emph{A} only in solid-red,
    edges in \emph{B} only in dashed-blue,
    and edges in both in gray.
    }
    \Description{Differences in the Causal Program Dependence Models for word count created using the four different partitions of the test suite}
    \label{fig:wc-ts}
    \end{figure*}

\subsubsection{Program Comprehension with Quantifiable Dependence}
 
Filtering out edges based on their direct dependences reveals patterns that relate 
to features in the source code. Consider Figure~\ref{fig:cpdm-8} that only 
shows edges with strong direct dependence. Nodes connected in such edges often 
correspond to features that are executed in all executions. For example, 
nodes \node{5}, \node{15}, and \node{16}, that are connected in 
Figure~\ref{fig:cpdm-8}, collectively implement the function of reading a 
character from the input file. Similarly, nodes \node{1} and \node{7} increase 
the character count in each iteration. Nodes \node{2}, \node{8}, and \node{9} 
increase the line count, while \node{10}, \node{13}, and \node{14} determine 
if the current character is part of a word.

Figure~\ref{fig:cpdm-28} shows the edges with intermediate direct dependence 
values between 0.2 and 0.8. Compared to Figure~\ref{fig:cpdm-8}, 
the connected nodes often reflect features that are only occasionally executed.
For example, the word counting feature is only executed when there is at least 
one non-alphabet character in the input. Since seven out of 15 tests (three 
single character tests and four single word tests) do not have non-alphabet 
characters, the nodes implementing word counting feature consequently have small direct 
dependences for the whole test suite. Such dependences (denoted \de{from}{to}) 
include \de{\node{20}}{\node{10}} and \de{\node{14}}{\node{11}}. 
Even a feature that is 
executed in every execution may have a small direct dependence. For example, 
the character read at node \node{16} affects the predicate at node \node{8}, 
which in turn checks whether the character is a newline or not. Since among 
many characters, only newline character makes the difference, their causal 
relationship is not strong. 

We posit that the capability to focus on different bands of direct dependences can 
help program comprehension. Despite being a small program of only about 40 
lines, its PDG with equally weighted edges makes it difficult to understand 
the program semantics. To highlight the differences between PDG and CPDM, we 
discuss a few edges that are not found in the CPDM and why there are good, 
semantically meaningful reasons for this.
 
\begin{itemize}[leftmargin=*]
\item \de{\node{5}}{all nodes control dependent on \node{5} except \node{15}}: 
\node{5} and \node{15} are the predicates of the main loop. Therefore, all 
nodes within the loop are control dependent on the two. Since \node{15} is in 
the loop, if \node{5} changes, \node{15} also changes (along with all other 
nodes in the loop). In this case \node{5} is considered not as a direct parent of any of 
the other nodes in the loop and, consequently, CPDM produces the edge 
\de{\node{5}}{\node{15}} while the PDG has many edges going out from \node{5}.

\item \de{\node{15}}{\node{8}, \node{10}, \node{17}}: \node{8}, \node{10}, and \node{17} are control dependent on \node{15}. However, they are also data dependent on \node{16}, (\node{19}, \node{20}), or \node{16}, respectively, which in turn depend on \node{15}. If \node{15} changes, \node{16}, \node{19}, and \node{20} changes before \node{8}, \node{10}, and \node{17}. Thus, the corresponding PDG edges are not in CPDM.
    
\end{itemize}

Let us now consider the relative strengths of dependence represented by the edge weights in CPDM. We highlight the following:

\begin{itemize}[leftmargin=*]

\item \de{\node{19}}{\node{10}} $>$ \de{\node{20}}{\node{10}}: In function \texttt{isLetter}, \node{19} and \node{20} are the nodes that express whether the character is an alphabetic character or not. Since seven out of the 15 tests do not have a non-alphabet character, the effect on \node{10}, which receives a return value of \texttt{isLetter}, of \node{19} is more significant than that of \node{20}. It is important to keep in mind that the CPDM takes into account if a node's behavior changes during execution, but not the number of different values it produces. Thus it is the \emph{number} of tests cases and not their length that is important here.

\item \de{\node{4}}{\node{11}} $>$ \de{\node{13}}{\node{11}} $>$ \de{\node{14}}
{\node{11}}: \node{4}, \node{13}, and \node{14} all correspond to the variable 
\texttt{inword}. \node{4} has the highest direct dependence on \node{11} because 
it affects whenever there exist more than zero alphabet characters. \node{13} 
and \node{14} affect \node{11} if the input includes more than one alphabet 
character. Because there are test cases that do not have any non-alphabet 
characters, \node{13} affects \node{11} more than \node{14}.

\end{itemize}
  
\subsubsection{Impact of Test Suites on CPDM}

Let us consider the impact of building the CDPM for \wc using the four 
categories of its test suites: \emph{onechar}, \emph{oneword}, \emph{oneline}, 
and \emph{multiline}. Figure~\ref{fig:wc-ts} shows \emph{differences} in the 
CPDMs constructed using different categories. The structure of CPDM changes according to the used category. In a figure labeled $A - B$, solid red edges are found only in $A$, dashed blue edges are found only in $B$, and gray edges are found in both $A$ and $B$. We highlight some of the findings below.

\begin{itemize}[leftmargin=*]
    
\item One character versus multiple characters (Figure~\ref{fig:word-char}): 
\node{15}, \node{16}, and \node{13} affect others only when there are more 
than one characters in the input. Therefore, there are outgoing edges from 
these in the CPDM of \emph{oneword} but not in \emph{onechar}.
Instead, the direct predecessor nodes of \node{15}, \node{16}, and \node{13}
(\node{5}, \node{6}, and \node{10}) are linked to the nodes affected by 
\node{15}, \node{16}, and \node{13} in \emph{onechar}.

\item One word versus multiple words (Figure~\ref{fig:line-word}): The 
sequence of edges \{\node{18}\} $\rightarrow$ \{\node{20}\} $\rightarrow$ 
\{\node{10}\} $\rightarrow$ \de{\node{14}}{\node{11}} is introduced in the 
CPDM created from \emph{oneline}: the model reflects the information 
flow from reading a non-alphabet character to the increment of the word 
counter. Comparison with Figure~\ref{fig:word-char} shows that the direct 
dependence of \de{\node{4}}{\node{11}} and \de{\node{13}}{\node{11}} have 
decreased due to the advent of a new node, \node{14}, affecting \node{11}. 
Note that, unlike in Figure~\ref{fig:cpdm-28}, the direct dependence of 
\de{\node{19}}{\node{10}} and \de{\node{20}}{\node{10}} is similar. This is 
because every test in \emph{oneline} contains both alphabet and non-alphabet
characters.

\item One line versus multiple lines (Figure~\ref{fig:m-1line}): With a 
newline character in the input, \de{\node{15}}{\node{8}} disappears, while 
\de{\node{16}}{\node{8}}, \de{\node{8}}{\node{9}}, and \de{\node{2}}{\node{9}}
appear: the new edges represent counting of lines. Furthermore, using \emph{multiline} makes the effect of \node{16} on \node{8} larger than that in 
Figure~\ref{fig:cpdm-28}, because every test case contains a newline character.
\end{itemize}

\noindent \textbf{Answer to RQ1:} The numerical causality associated with each 
dependence tells us the strength of the causal connection between the two 
program elements involved. Our investigation shows that by exploiting these 
weights, CPDM can effectively focus a developer on dependencies of greater 
importance (those with higher causality). This focus provides a better 
understanding of the program.

\begin{table}[ht]
    \caption{Acc@n for CDFL ($N_{mpn} = 20$, median), SBFL, Dicing, and DS. Avg., LO, Max, and Min are different tiebreakers.}
    \small
    \label{tbl:accn}
    \resizebox{\columnwidth}{!}{
    \begin{tabular}{rrrrrrrrrrrrrr}
    \hline
    \multicolumn{1}{c}{\multirow{2}{*}{$n$}} & \multicolumn{1}{c}{\multirow{2}{*}{CDFL}} & \multicolumn{4}{c}{SBFL}                                                                                    & \multicolumn{4}{c}{DS}                                                                                & \multicolumn{4}{c}{Dicing}                                                                            \\
    \multicolumn{1}{c}{}                     & \multicolumn{1}{c}{}                      & \multicolumn{1}{c}{Avg.} & \multicolumn{1}{c}{LO}       & \multicolumn{1}{c}{Max} & \multicolumn{1}{c}{Min} & \multicolumn{1}{c}{Avg.} & \multicolumn{1}{c}{LO} & \multicolumn{1}{c}{Max} & \multicolumn{1}{c}{Min} & \multicolumn{1}{c}{Avg.} & \multicolumn{1}{c}{LO} & \multicolumn{1}{c}{Max} & \multicolumn{1}{c}{Min} \\ \hline
    1                                        & 8                                         & 3                        & \textbf{10} & 3                       & 21                      & 0                        & 0                      & 0                       & 92                      & 3                        & 7                      & 3                       & 9                       \\
    3                                        & \textbf{19}              & 11                       & 14                           & 9                       & 26                      & 0                        & 3                      & 0                       & 92                      & 7                        & 7                      & 7                       & 9                       \\
    5                                        & \textbf{27}              & 17                       & 18                           & 12                      & 41                      & 0                        & 6                      & 0                       & 92                      & 7                        & 7                      & 7                       & 9                       \\
    10                                       & \textbf{43}              & 31                       & 40                           & 27                      & 59                      & 2                        & 12                     & 0                       & 92                      & 7                        & 8                      & 7                       & 9                       \\ \hline
    \end{tabular}}
    \vspace{-.7em}
    \end{table}

\subsection{Performance of CDFL}
\label{sec:rq2}

Here, we report the accuracy of CDFL in comparison to SBFL and other 
slice-based FL techniques.

\subsubsection{Overall Performance of CDFL}

Table~\ref{tbl:accn} contains $acc@n$ metrics from the four studied FL 
techniques: CDFL, SBFL, Dynamic Slicing, and Dicing. The CDFL results are 
median values from ten trials, each based on CPDA computed with 20 sampled 
mutants (i.e., $N_{mpn} = 20$). CDFL and SBFL outperform the slicing-based 
techniques for all values of $n$. CDFL performs better than SBFL rankings 
computed with the average tiebreaker: it can locate 166.6\%, 72.7\%, 58.8\%, 
and 38.7\% more faults in the top 1, 3, 5, and 10 places, respectively. 
Compared to SBFL with the line order tiebreaker, CDFL performs better in all 
cases except for $acc@1$. Note that SBFL, which is known to require diverse 
test cases for better performance~\cite{Steimann:2013sf}, suffers from the use 
of coverage adequate test suites that typically contain only one or two 
failing test cases. CDFL achieves better performance using the same test suite.

Since DS and Dicing use equally weighted dependence, all of their results are 
tied by definition. In most cases, the faulty line is executed by both passing 
and failing tests, which adversely affect the performance of dicing: among 92 
faults, only nine faulty lines are found in the dice, all tied. SBFL is also 
vulnerable to ties, which can be seen from the results from minimum and 
maximum tie-breakers. On average, the number of tied program elements that are 
tied with the root cause is 20 for SBFL, and 95 for DS. In comparison, CDFL 
does not produce any ties.

\subsubsection{Impact of Mutant Sampling on CDFL}

To investigate the effect of the number of sampled mutants for each node, we 
vary the value of $N_{mpn}$ from 2 to 20 by intervals of two and repeat the 
CDFL analysis for each configuration. Our preliminary evaluation with 
$N_{mpn} \in \{30, 40, 50\}$ did not improve the accuracy of CDFL 
significantly beyond that obtained with $N_{mpn} = 20$. Note that we increase 
$N_{mpn}$ by adding more samples (i.e., the set of samples with a larger 
$N_{mpn}$ includes the set with a smaller $N_{mpn}$): this is to avoid sampling 
bias when considering the effect of $N_{mpn}$. 

\begin{figure}[ht]
\centering
\includegraphics[width=\columnwidth,trim={70 30 105 50},clip]{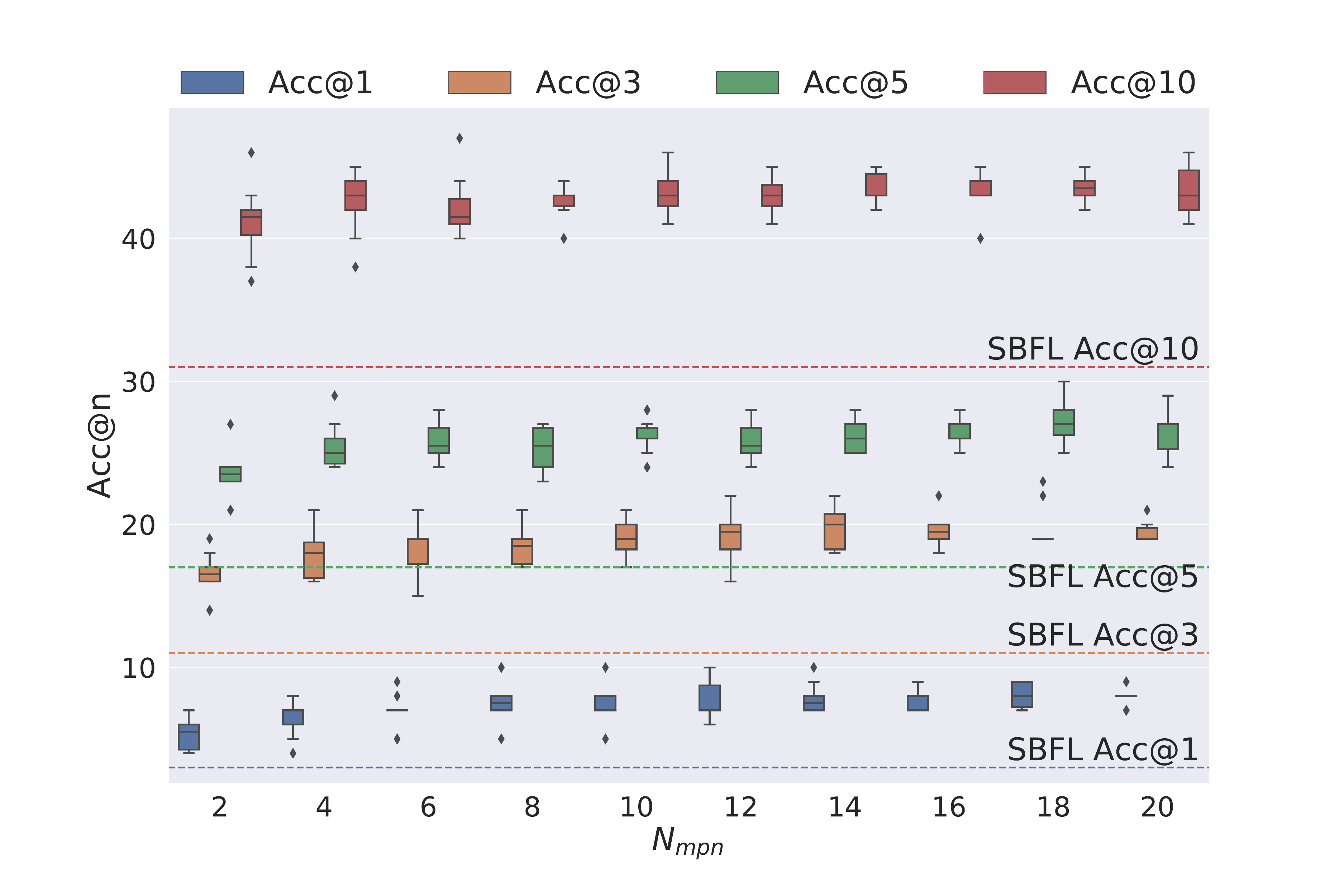}
\caption{Acc@n of CDFL with different $N_{mpn}$ and SBFL using the average tiebreaker}
\Description{Accuracy at n of Causal Dependence Fault Localization with different 
number of mutation samples per node and Spectra based Fault Localization using 
the average tiebreaker}
\vspace{-1.4em}
\label{fig:accn}
\end{figure}

Figure~\ref{fig:accn} shows boxplots of CDFL results from ten trials, obtained 
with varying $N_{mpn}$, in comparison to the results of SBFL, which are shown 
by the dashed horizontal lines. The boxplots initially show an increasing 
trend as $N_{mpn}$ increases. However, the trend tends to stabilize after 
$N_{mpn} = 10$: the difference between median of $acc@n$ at $N_{mpn} = 10$ and 
$N_{mpn} = 20$ is less than, or equal to, one.  The result suggests that, for 
the studied programs, ten mutation samples for each node is sufficient for 
fault localization.

\begin{table*}[ht]
\caption{Acc@n for CDFL ($N_{mpn} = 20$, median) and SBFL (avg. tie breaker) for each subject program}
\label{tbl:accn-prog}
\scalebox{0.8}{
\begin{tabular}{r|r|rrrr|rrrr|rrrr|rrrr|rrrr|rrrr|rrrr}
\toprule
\multicolumn{2}{r|}{Program (\# bugs)}    & \multicolumn{4}{c|}{\tcas (34)} & \multicolumn{4}{c|}{\sch (4)} & \multicolumn{4}{c|}{\scht (3)} & \multicolumn{4}{c|}{\tot (18)} & \multicolumn{4}{c|}{\prt (3)} & \multicolumn{4}{c|}{\prtt (6)} & \multicolumn{4}{c}{\rep (24)} \\
\multicolumn{2}{r|}{$n$} & 1     & 3      & 5     & 10    & 1      & 3      & 5      & 10     & 1      & 3      & 5      & 10      & 1      & 3     & 5       & 10     & 1       & 3       & 5      & 10      & 1       & 3       & 5       & 10      & 1     & 3     & 5       & 10     \\ \midrule
\multirow{2}{*}{Acc@n} & CDFL       & \textbf{4}     & \textbf{10}     & \textbf{14}    & \textbf{23}    & 0      & \textbf{1}      & 1      & \textbf{2}      & \textbf{1}      & \textbf{1}      & \textbf{1}      & \textbf{1}       & 0      & \textbf{1}     & \textbf{2.5}     & 3      & 0       & \textbf{1}       & 1      & \textbf{2}       & 1       & 2       & 2       & 3       & 2     & 3     & 4.5     & 9      \\
                       & SBFL       & 0     & 3      & 7     & 10    & 0      & 0      & 1      & 1      & 0      & 0      & 0      & 0       & 0      & 0     & 0       & 3      & 0       & 0       & 1      & 1       & 1       & 2       & 2       & 3       & 2     & \textbf{6}     & \textbf{6}       & \textbf{13}    \\
\bottomrule
\end{tabular}
}
\end{table*}

\subsubsection{Impact of Causal Structure on CDFL}

Let us consider the impact that causal structure discovery has on the accuracy 
of CDFL. Table~\ref{tbl:accn-prog} presents the per-subject breakdown of 
$acc@n$ results from CDFL and SBFL. CDFL produces significantly higher $acc@n$ 
for \tcas, while performing much worse than SBFL for \rep. For other subjects, 
both techniques perform similarly well.

We posit that CDFL performs better for \tcas due to its simpler control-flow 
structure. \tcas is an aircraft collision avoidance system that calculates the 
altitude separation~\cite{Hutchins:1994aa}. The whole program is a sequence of 
logical formulas for the calculation of a single value, which requires minimal 
branching. Consequently, many computation steps are performed by both passing 
and failing test executions, presenting a challenge for SBFL. In comparison, 
CDFL considers both control and data dependency from the node trajectories, 
resulting in better performance.

In contrast, for many of the \rep faults for which CDFL performs poorly, we 
find that the causal structure discovery is inadequate. By definition, a 
causal structure is a DAG, and cannot properly represent the cyclic dependence 
in a program. Breaking these cycles can lead to an incorrect set of Markovian 
parents, which in turn results in incorrect causal dependences. \rep takes a 
a regex pattern, a substitution string, and a target text: it matches the 
pattern in the target text and replaces the match with the substitution 
string. During its operation, \rep constructs lots of cyclic dependence 
relations between program elements. Such a large number of the cyclic 
dependencies cause CPDA to struggle to model the causal dependence accurately.

\noindent\textbf{Answer to RQ2:} CDFL can significantly outperform other 
slicing based fault localisation techniques. With sufficient mutant sampling, 
it can outperform SBFL and does not suffer from ties.

\section{Discussion \& Future Work}
\label{sec:discussion}

We currently use the definition of Markov parents directly in our algorithm to 
calculate causal dependences and the strengths of the causal connections in the 
CPDM. This is motivated by the fact that methods for causal discovery 
typically focus on observational data only, while we intervene in the program 
execution and thus have interventional data. Furthermore, our algorithm starts 
from binarized, indicator variables rather than from the raw observations. 
This is natural when, at least on the level of a high-level programming 
language, variables can contain complex structures such as trees.

However, in recent years there has been advances in learning causal structure 
also from interventional data and there are even algorithms for calculating 
\textit{optimal} models with cycles~\cite{Rantanen:2020learning}. Future work 
should investigate how to apply these advances and tools in our setting. One 
challenge will be to bridge the gap from a high-level program with complex 
values in the nodes to methods that assume numerical data only. On the other 
hand, there might be several benefits, such as being able to handle cyclic dependences, as well as faster, more scalable structure learning~\cite{Ramsey:2017million}.

\section{Threats to Validity}
\label{sec:threats}

Using a limited set of program inputs and approximating the dependence is a typical 
internal threat to any dynamic analysis, and thus also to CPDA. CPDA gets less 
effect from a small number of inputs through various mutations compared to 
dynamic slicing. RQ1 analyzes the effect of input by observing changes in CPDM 
when using different tests. For RQ2, we selected the tests based on the 
statement coverage criteria and used the same tests for other FL techniques to 
mitigate the threat. The sampling of mutants for CPDA poses another threat to the 
internal validity. To mitigate it,
we sample a 
sufficiently large number of 100 mutants for RQ1, and conduct 10 independent 
trials for RQ2. 

The use of the Siemens suite poses a threat to external validity. While we 
cannot claim that the result of RQ2 generalize,
the Siemens suite is widely used in fault localization studies and enable comparison. 
We rely on 
a qualitative analysis for RQ1, and a widely studied metric ($acc@n$) for RQ2, 
to minimize the threat.

\section{Related Work}
\label{sec:relatedwork}

This paper discusses two mainstream software engineering tasks: program 
dependence analysis and fault localization.

\subsection{Program Dependence Analysis}

Static analysis attempts to uncover facts about a program that apply to any 
possible execution. It is therefore necessarily conservative and consequently 
often produces many false positives. Static dependence analysis is often used to produce a Program Dependence Graph (PDG), which was first use in compiler optimization and parallelization~\cite{
ferrante:graph}, and has subsequently found many uses~\cite{horwitz:the-use} including program slicing~\cite{
horwitz:the-use, horwitz:interprocedural-toplas}.

Dynamic dependence analysis incorporates one of more program inputs. A simple 
example is early dynamic slicing algorithms that computed a static slice of 
the PDG and then remove edges that are not executed~\cite{Agrawal:1990aa}.
Probabilistic Program Dependence Graph (PPDG) augments the PDG with a 
set of abstract states at each node that enable the use of probabilistic 
reasoning to analyze program behavior~\cite{Baah:2010aa}.
Later, the Bayesian Network based Program Dependence Graph (BNPDG) 
augments the PDG with conditional probabilities to relate the
state of a node to that of its parents~\cite{Yu:2017ab}.

In comparison, CPDM is not tied to PDG, freeing us from having to solve hard 
data-flow problems, such as pointer analysis. Instead, we extract the 
dependence structure from observations of program executions. We expect to 
further exploit the advances in the fields of causal inference and causal 
discovery to refine our approach. In contrast, PPDG~\cite{Baah:2010aa} and 
BNPDG~\cite{Yu:2017ab} are tied to the lower levels of the causal hierarchy 
as introduced by Pearl~\cite{Pearl:2019seven} as they are based on the associative conditional probability between program states.

\subsection{Fault Localization}

Fault Localization (FL) aims to identify parts of the program source code that 
are likely to be the root cause of the observed test failures: typically FL 
techniques rank program elements by their relative likelihood of being the 
root cause~\cite{Wong:2016aa}. One of the most widely studied FL techniques is
Spectrum Based Fault Localization (SBFL), a dynamic approach that ranks 
program elements based on their suspiciousness, which is computed from test 
coverage and outcomes~\cite{Steimann:2013sf}. SBFL has been widely studied 
both as an independent technique~\cite{Abreu:2007aa,Xie:2013uq,Naish:2011fk} 
and in a hybridization with other FL input features and techniques~\cite{
Sohn:2017aa,B.-Le:2016yu,Li2019aa,Le:2015aa}. However, it tends to produce 
many ties, as some program elements can share the same test coverage and 
outcome. Being based on coverage, SBFL also suffers from Coincidental 
Correctness (CC), i.e., passing executions that cover faulty elements~\cite{Masri:2010ph}.

Several existing work utilize the causal inference to fault localization. 
Baah et al.~\cite{Baah:2010fj,Baah:2011ab} use a linear model to capture the 
causal effect from coverage of program elements to test outcomes. Gore et 
al.~\cite{Gore:2012aa} and Shu et al.~\cite{Shu:2013aa} apply a 
similar linear regression approach to predicate values and method level 
coverage, respectively. In comparison, CDFL focuses on general dependence 
relations from CPDA instead of coverage, and therefore is less affected by CC.

\section{Conclusion}
\label{sec:conclusion}

We propose CPDA that measures strengths of 
dependence between program elements by modeling their causal relationship. 
Applying causal inference on observational data, instead of using static 
analysis, frees dependence analysis from the burden of pointer analysis and 
avoids producing large number of equally weighted dependence relations. 
Our evaluation of CPDM, a graph representation of a program causal relations, 
shows that CPDA can cluster semantically related 
program elements, based on meaningful differences in the strengths of program 
dependence. We also show the utility of continuously quantifiable program
dependence by using it to create a novel fault localization method (CDFL). 
Our empirical evaluation shows that CDFL can outperform other FL techniques 
even with limited test suites. Future work will consider more advanced causal 
modeling with an emphasis on cyclic structures, as well as other applications 
of CPDA.

\bibliographystyle{ACM-Reference-Format}
\bibliography{ref}

\newpage

\appendix

\section{Direct effect in CPDA}
  \label{sec:appendix}

  In this section, we reduce the formula of Definition~\ref{def:DD_def}.

  \begin{theorem}
    For a subset $X$ of random nodes,
    \small
    \begin{equation*}
      \label{equ:CE_set}
      P(s_1, \cdots, s_n \mid \hat{x}) = \begin{cases}
        \prod_{i \mid S_i \notin X} P(s_i \mid pa_i) & \parbox{.35\columnwidth}{for $s_1, \cdots, s_n$ consistent with $x$} \\
        0 & \text{otherwise}
      \end{cases}\,.
    \end{equation*}
    \normalsize
  \end{theorem}

  \vspace{1em}

  \begin{theorem}
    Let $G = (V, E)$ is a causal structure. Given two disjoint sets of nodes, $X,Y \in V$,
    \small
    \begin{equation}
      P(y \mid \hat{x}) = \sum_{\left(s_{i_1}, \cdots, s_{i_m}\right) \mid S_i \notin Y} P(s_1, \cdots, s_n \mid \hat{x})\,.
    \end{equation}
    \normalsize
  \end{theorem}

  \vspace{1em}

  \begin{corollary}
    \label{def:CE_coro1}
    \small
    \begin{align*}
      P(y \mid \hat{x}') &= \sum_{\left(s_{i_1}, \cdots, s_{i_m}\right) \mid S_i \notin Y} P(s_1, \cdots, s_n \mid \hat{x}') \\
      &= \sum_{\left(s_{p_1}, \cdots, s_{p_q}\right) \mid S_p \notin Y \cup X} 
      \left[\sum_{\left(s_{r_1}, \cdots, s_{r_s}\right) \mid S_r \notin Y \cup X^c} P(s_1, \cdots, s_n \mid \hat{x}')\right] \\
      &= \sum_{\left(s_{p_1}, \cdots, s_{p_q}\right) \mid S_p \notin Y \cup X} 
      \left[\sum_{s_{r_1}} \cdots \sum_{s_{r_{s-1}}} \sum_{s_{r_s}} P(s_1, \cdots, s_n \mid \hat{x}')\right]\,, \\
      \\ &\text{where $\left\{s_{r_1}, \cdots, s_{r_s}\right\} \in X \cap Y^c$; since Theorem \ref{equ:CE_set},}\\\\
      &= \sum_{\left(s_{p_1}, \cdots, s_{p_q}\right) \mid S_p \notin Y \cup X} 
      \left[\sum_{s_{r_1}} \cdots \sum_{s_{r_{s-1}}} P(s_1, \cdots, s_{r_s}', \cdots, s_n \mid \hat{x}')\right] \\
      = \cdots &= \sum_{\left(s_{p_1}, \cdots, s_{p_q}\right) \mid S_p \notin Y \cup X} P(s_1, \cdots, s_n \mid \hat{x}')
    \end{align*}
    \normalsize
  \end{corollary}

  \vspace{1em}

  \begin{corollary}
    \label{def:CE_coro2}
    For a statement $Y \in V$, using Corollary~\ref{def:CE_coro1},
    \small
    \begin{align*}
      P(y \mid \hat{pa}_Y) 
      &= \sum_{\left(s_{p_1}, \cdots, s_{p_q}\right) \mid S_p \notin \left\{Y\right\} \cup PA_Y} P(s_1, \cdots, s_n \mid \hat{pa}_Y) \\
      &= \sum_{\left(s_{p_1}, \cdots, s_{p_q}\right) \mid S_p \notin \left\{Y\right\} \cup PA_Y} \left[ \prod_{j \mid S_j \notin PA_Y} P(s_j \mid pa_j)\right] \quad \text{(by Theorem~\ref{equ:CE_set})}\\
      &= P(y \mid pa_Y) \sum_{\left(s_{p_1}, \cdots, s_{p_q}\right) \mid S_p \notin \left\{Y\right\} \cup PA_Y} \left[ \prod_{j \mid S_j \notin \left\{Y\right\} \cup PA_Y} P(s_j \mid pa_j)\right] \\
      &= P(y \mid pa_Y)\,,
    \end{align*}
    since $\sum_{\left(s_{p_1}, \cdots, s_{p_q}\right) \mid S_p \notin \left\{Y\right\} \cup PA_Y} \left[ \prod_{j \mid S_j \notin \left\{Y\right\} \cup PA_Y} P(s_j \mid pa_j)\right] = 1$.
    \normalsize
  \end{corollary}

  \vspace{1em}

  \begin{corollary}
    \label{def:BD_coro}
    \small
    \begin{align*}
      & \mathit{DD}(S_i, S_j) \\
      & = \frac{1}{|I|}\sum_{i \in I} \mathit{NDE}_{O_i, S_i: 0 \rightarrow 1}(S_j) \\
      & = \frac{1}{|I|}\sum_{i \in I}
      \sum_{z} 
        \left[
          \left\{
            E_{O_i} \left(S_j \mid do(S_i = 1, z)\right) 
            - E_{O_i} \left(S_j \mid do(S_i = 0, z)\right)
          \right\} \right.\\
          &\qquad\qquad\qquad\left.
          \times \  P_{O_i}\left(z \mid do(S_i = 0)\right) 
        \right] \\
      & = \frac{1}{|I|}\sum_{i \in I}
      \sum_{z} 
        \left[
          \left\{
            P_{O_i} \left(S_j = 1 \mid do(S_i = 1, z)\right) 
            - P_{O_i} \left(S_j = 1 \mid do(S_i = 0, z)\right)
          \right\} \right.\\
          &\qquad\qquad\qquad\left.
          \times \  P_{O_i}\left(z \mid do(S_i = 0)\right) 
        \right]\,, \\
      \\ & \text{where $Z = PA_j \backslash S_i$. Since $\left\{S_i\right\} \cup Z = PA_j$, by Corollary~\ref{def:CE_coro2},} \\\\
      & = \frac{1}{|I|}\sum_{i \in I}
      \sum_{z} 
        \left[
          \left\{
            P_{O_i} \left(S_j = 1 \mid S_i = 1, z\right) 
            - P_{O_i} \left(S_j = 1 \mid S_i = 0, z\right)
          \right\} \right.\\
          &\qquad\qquad\qquad\left.
          \times \  P_{O_i}\left(z \mid do(S_i = 0)\right) 
        \right]\,.
    \end{align*}
    \normalsize
    Here, by Corollary~\ref{def:causal_effect},
    \small
    \begin{align*}
      P_{O_i}(z \mid do(S_i = 0)) = \sum_{pa_i} P_{O_i}(z \mid S_i = 0, pa_i) P_{O_i}(pa_i).
    \end{align*}
    \normalsize
  \end{corollary}

\end{document}